\documentclass[pra,twocolumn,superscriptaddress,aps]{revtex4}
\usepackage{amsfonts}
\usepackage{amssymb,latexsym,amsmath}
\usepackage{times}
\usepackage[caption=false]{subfig}
\usepackage[usenames]{color}
\usepackage{graphicx,tikz}
\usepackage{fancyvrb}

\begin{document}

\title{Evaluating the Robustness of Rogue Waves Under Perturbations}

\author{C. B. Ward}
\affiliation{Department of Mathematics and Statistics, University of
	Massachusetts, Amherst MA 01003-4515, USA}

\author{P. G. Kevrekidis}
\affiliation{Department of Mathematics and Statistics, University of
	Massachusetts, Amherst MA 01003-4515, USA}

\author{N. Whitaker}
\affiliation{Department of Mathematics and Statistics, University of
	Massachusetts, Amherst MA 01003-4515, USA}
	
\begin{abstract}
  Rogue waves, and their periodic 
  counterparts, have been shown to exist in a number of integrable
  models. However, relatively little is known about the existence of these objects in models where an exact formula is unattainable. In this work, we develop a novel numerical perspective towards identifying such
  states as localized solutions in space-time. Importantly, we
  illustrate that this
  methodology in addition to benchmarking known solutions (and
  confirming their numerical propagation under controllable error)
  enables the continuation of such solutions over parametric variations
  to non-integrable models. As a result, we can answer in the positive
  the question about the parametric robustness of Peregrine-like waveforms
and even of generalizations thereof on a cnoidal wave background.
\end{abstract}

\maketitle
        
\section{Introduction: Motivation \& Approach}

Over the past decade there has been a tremendous
explosion of interest towards the study of phenomena
associated with extreme wave events, the so-called rogue or freak waves.
This has largely been triggered by the experimental realization
(and associated control) of such waveform manifestations in a diverse
array of experiments that range from nonlinear
optics~\cite{opt1,opt2,opt3,opt4,opt5,laser} to
hydrodynamics~\cite{hydro,hydro2,hydro3} and from
plasmas~\cite{plasma} to superfluid helium~\cite{He}
and to parametrically driven capillary waves \cite{cap}.
The relevant advances have, by now, been discussed
in numerous books on the subject~\cite{k2a,k2b,k2c,k2d}
and have, in turn, triggered considerable progress
towards the theoretical study of such waves, which are now
contained to a large extent in a number of
reviews~\cite{yan_rev,solli2,onorato}.

Much of the relevant theoretical activity has revolved
around the feature that in integrable models such as the
nonlinear Schr{\"o}dinger (NLS) equation the machinery
of the inverse scattering transform (IST) can be utilized
to obtain some prototypical waveforms that are
natural candidates as rogue wave structures. Typical
examples include the Peregrine soliton (PS)~\cite{H_Peregrine},
the Kuznetsov~\cite{kuz},
Ma~\cite{ma} (KM) soliton, and the Akhmediev breather (AB)~\cite{akh}, among others
(see also the work of Dysthe and Trulsen~\cite{dt}).
Among them, perhaps the most well-known entity is the
Peregrine soliton which is algebraically
localized in both space and time; the PS, KM and AB are essentially
members of a parametric family of solutions where variation of a
suitable parameter moves between a spatially periodic solution
(AB), a localized one in both space and time (PS), and a periodic
in time solution (KM).

However, a key lingering question is whether departure from integrability
allows for the persistence of such rogue waves. To the best of our
understanding the attempts to address this issue have been quite
limited. For instance, in a special case example the work of~\cite{Ank1}
illustrated that under some realistic perturbations (such as third order
dispersion or self-steepening terms), a leading order perturbed variant
of the PS would persist. Similarly, a perturbed inverse scattering
approach has been used to consider the KM solution under dispersive
and dissipative perturbations~\cite{kgarnier}. Other authors have attempted to
argue on the basis of more general grounds~\cite{calinibook}, such
as the proximity of these solutions to chaotic states,
%of more integrablemodels to argue 
that they may persist. Nevertheless, it is clear that
these approaches each have some limitations (typically of finite
order considerations and of perturbative thus suggestive, yet
not conclusive treatment and the typical {\it loss} of integrability in
the presence of additional terms).

It thus remains a rather open question whether rogue wave structures
{\it persist} in the presence of ``generic'' perturbations and how
their profile may be modified as a result of these perturbations. The
unavailability of the integrable machinery of the IST which has been
used for the vast majority of results on the
subject~\cite{yan_rev,solli2,onorato} renders this
question even more dire. This is especially so in light of the fact that in
most cases that we are familiar with the integrable models like the
NLS are, at best, an idealized approximation of the true physical system.
Hence, if these solutions are to be relevant in realistic settings, their
persistence needs to be ensured. To address this question, numerical
techniques appear to be well suited: they are not limited in any way
by integrability considerations (on the contrary, they can
use the integrable limit as a useful starting point towards exploring
continuations to non-integrable variants). Furthermore, they
can provide a result accurate to the prescribed numerical tolerance
(and hence are not bound to ``leading order'' type considerations).
However, there is a nontrivial obstacle: the most appropriate way
to find localized solutions in these classes of models is via fixed
point iterations [be they Newton-type schemes,
  spectral renormalization schemes,
  or imaginary time variants~\cite{yangbook}]. Nevertheless, this
class of methodologies cannot be applied here, as the solution is
not stationary in time. Hence, if we are seeking a PS (a solution
that ``appears out of nowhere and disappears without a trace''~\cite{wandt}),
the proper way to think of the solution is as a localized one, a
{\it two-dimensional homoclinic orbit},
in space and time. Thus, we propose to consider
time as a {\it spatial variable} and to develop an iterative scheme
that identifies a localized solution in the $(x,t)-$plane. In the next
section we will present more about the method and the examples
(as well as benchmarks) of interest, while in the final section,
we will summarize our conclusions and pave some of the numerous
directions that open up for future study.

\section{Implementation: Benchmarks \& Results}

Motivated partly by recent explorations at the interface
of rogue waves and potentially collapsing dynamics~\cite{stathis}
(involving power law nonlinearities)
as well as by the relevance of perturbative terms (such as third
order dispersion (TOD)~\cite{Ank1}
in optics), we take as
our prototypical model example a two-parameter variation of the NLS:
\begin{equation} \label{eq1}
i\frac{\partial \psi}{\partial t} + \frac{1}{2}\frac{\partial^2 \psi}{\partial x^2}  + |\psi|^{2p} \psi - i\epsilon \frac{\partial^3 \psi}{\partial x^3}= 0.
\end{equation}
Nevertheless, we should highlight that our computational analysis is
by no means constrained to this particular choice. Indeed, it is expected to be of relevance
to a wide range of previously treated variants of the NLS model.

A relevant point to recall is  the modulation instability inherent
in the background of the PS.
As a result of this, finding rogue waves either
through time-integration methods (which are not particularly
  well-suited anyway, given that for arbitrary variants of NLS, it is
  not clear what initial data to use to obtain a PS)
or through fixed point iteration has been, in our experience,
especially difficult.
In that light, we have used 
a highly-efficient method, namely a variant of the
Newton-conjugate gradient method of~\cite{Yang}, originally designed for solitons on a zero background.

\begin{figure}[b]
	\subfloat[]{\includegraphics[width=.25\textwidth]{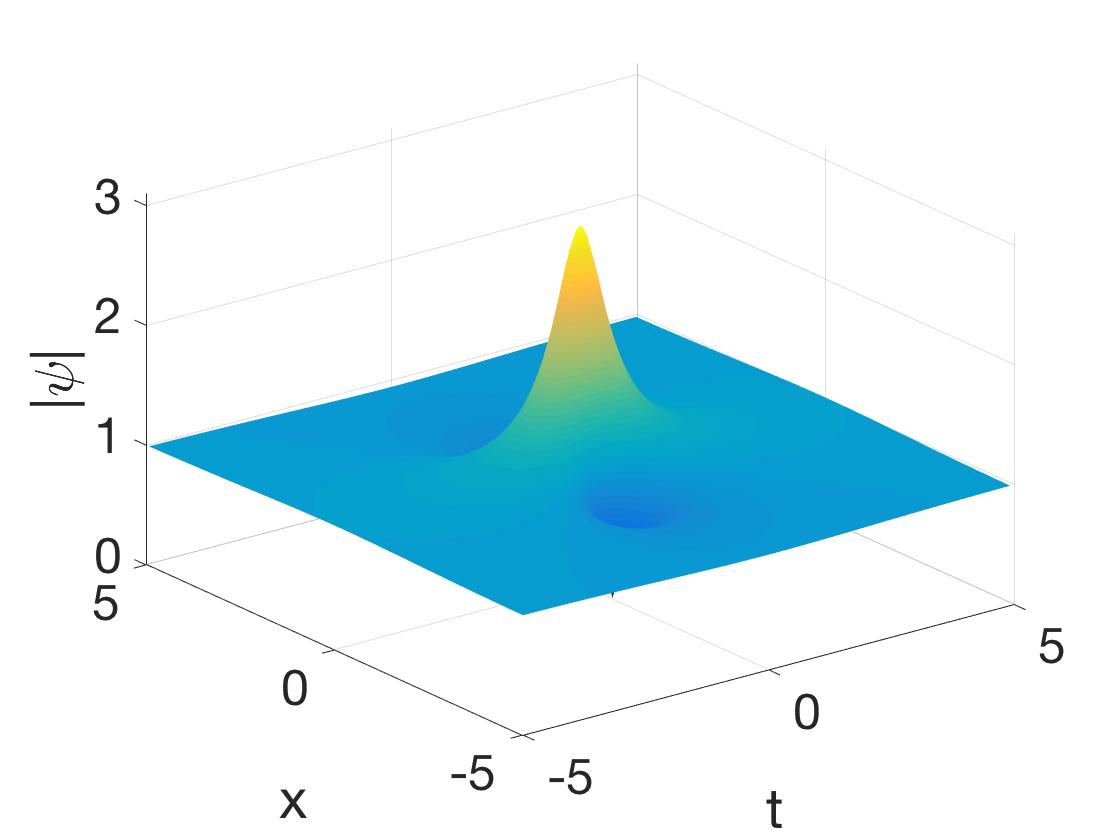}}
	\subfloat[]{\includegraphics[width=.25\textwidth]{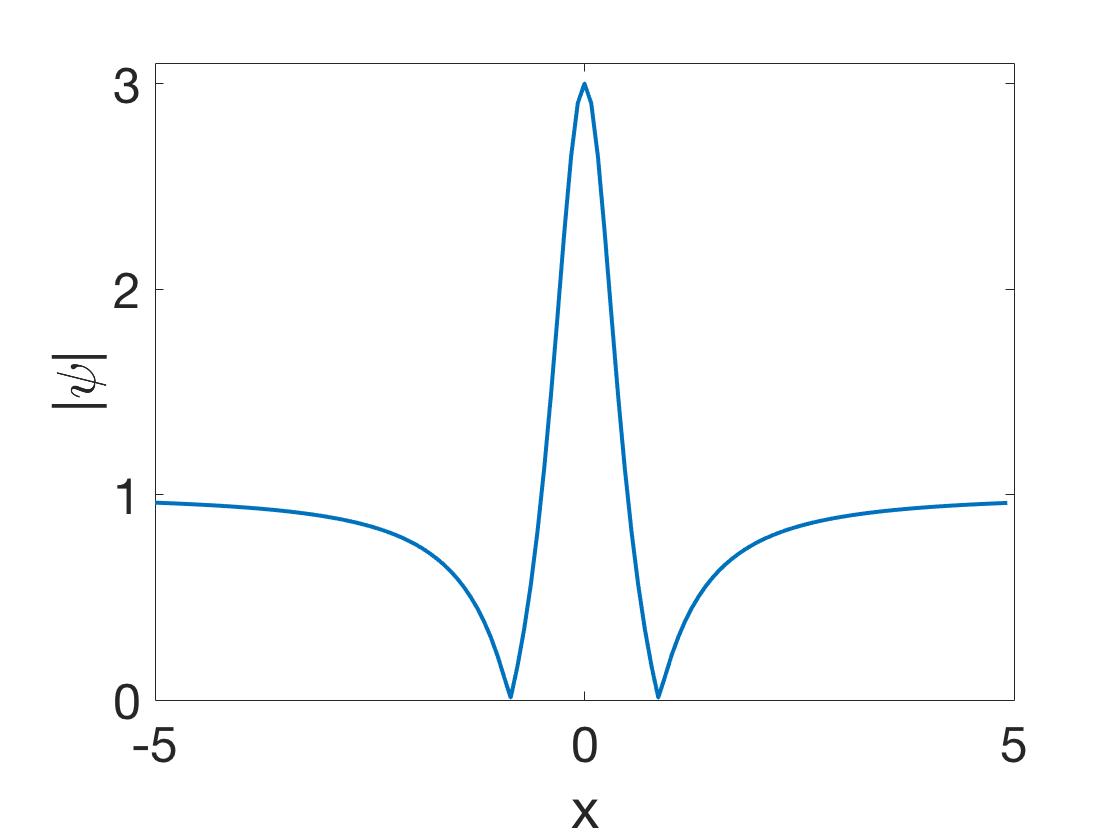}}
	
	\subfloat[]{\includegraphics[width=.25\textwidth]{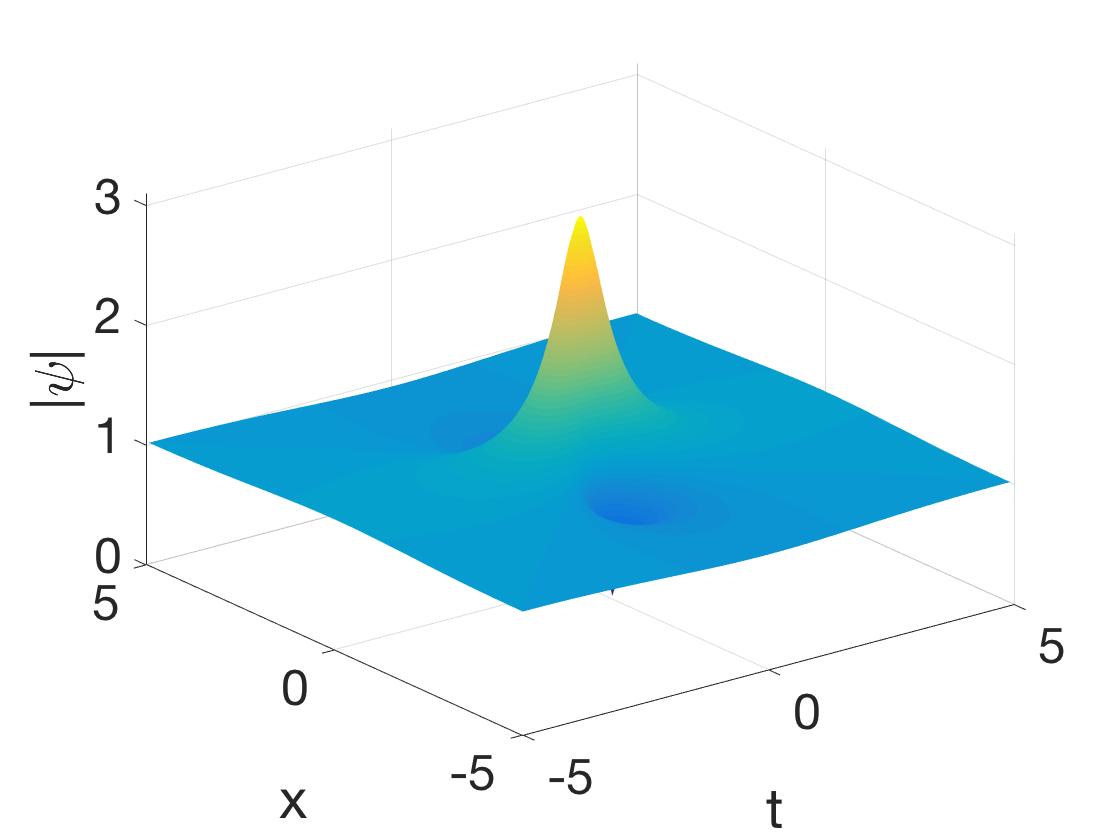}}
	\subfloat[]{\includegraphics[width=.25\textwidth]{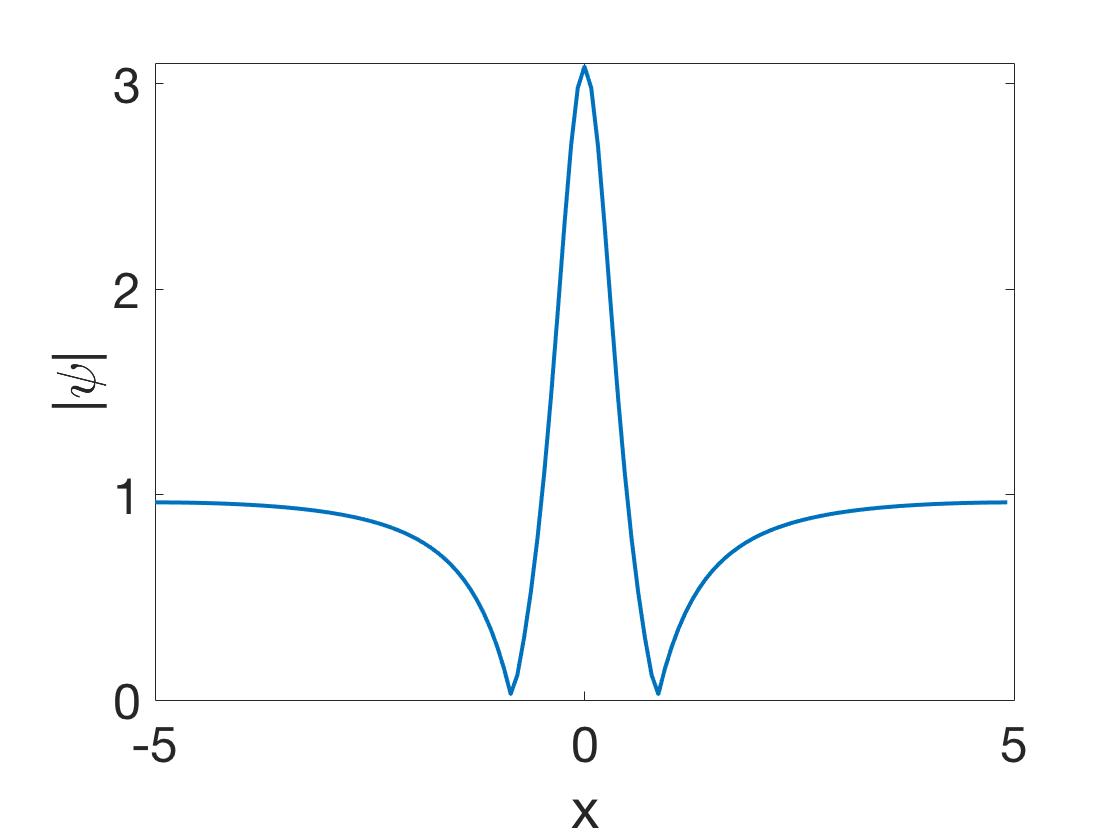}}
     \caption{Comparison of the Peregrine soliton (a) with the numerical solution (c). Time slice at $t=0$ of the exact solution (b) and numerical solution (d). Excellent agreement is seen between the two solutions.} \label{Fig1}
\end{figure}

For the NLS equation, our benchmark studies
show that the above Newton-CG method converges not only
to a good approximation of the PS but also to other families of
rogue waves. Fig.~\ref{Fig1} shows the exact Peregrine soliton (a) compared to that obtained by the numerical method (c). Fig.~\ref{Fig1}(b) and Fig.~\ref{Fig1}(d) show the spatial cross section at time $t=0$ for the analytical and numerical solutions, respectively. We see that despite the periodic boundary conditions the two solutions are nearly identical; in fact,
the pointwise error is on the order of $10^{-2}$.

\begin{figure}[t]
	\subfloat[]{\includegraphics[width=.25\textwidth]{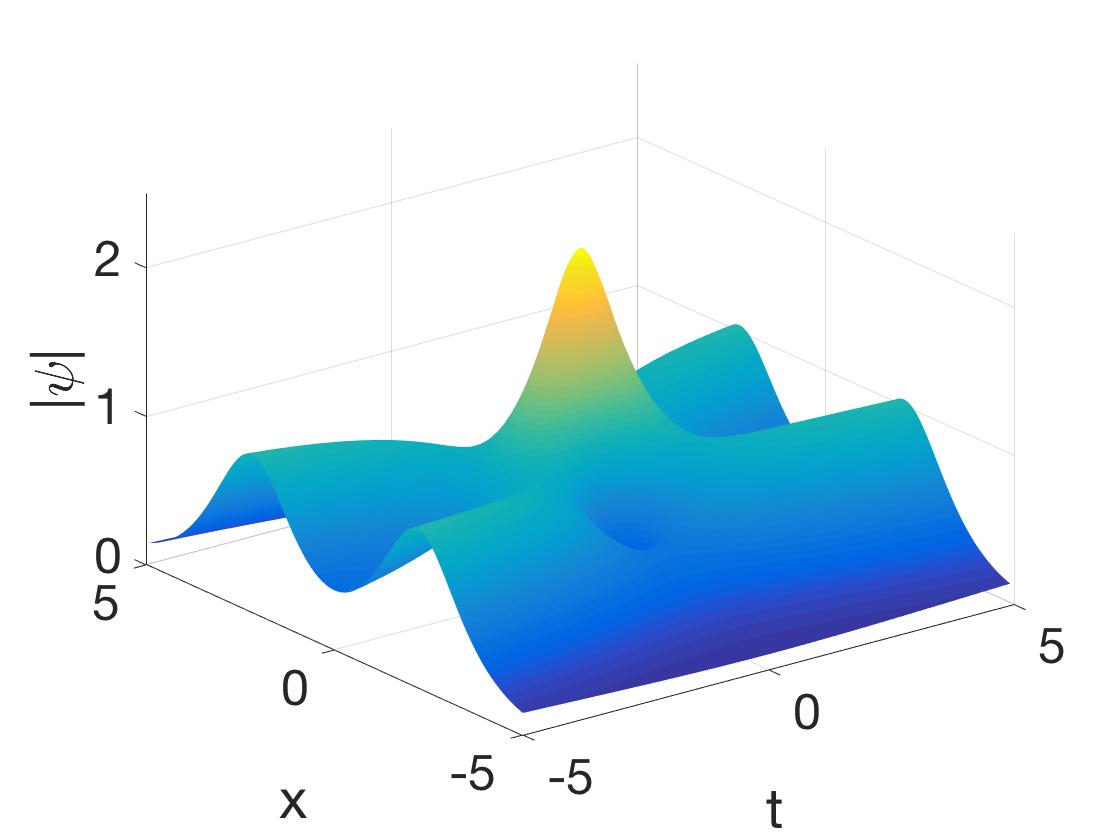}}
	\subfloat[]{\includegraphics[width=.25\textwidth]{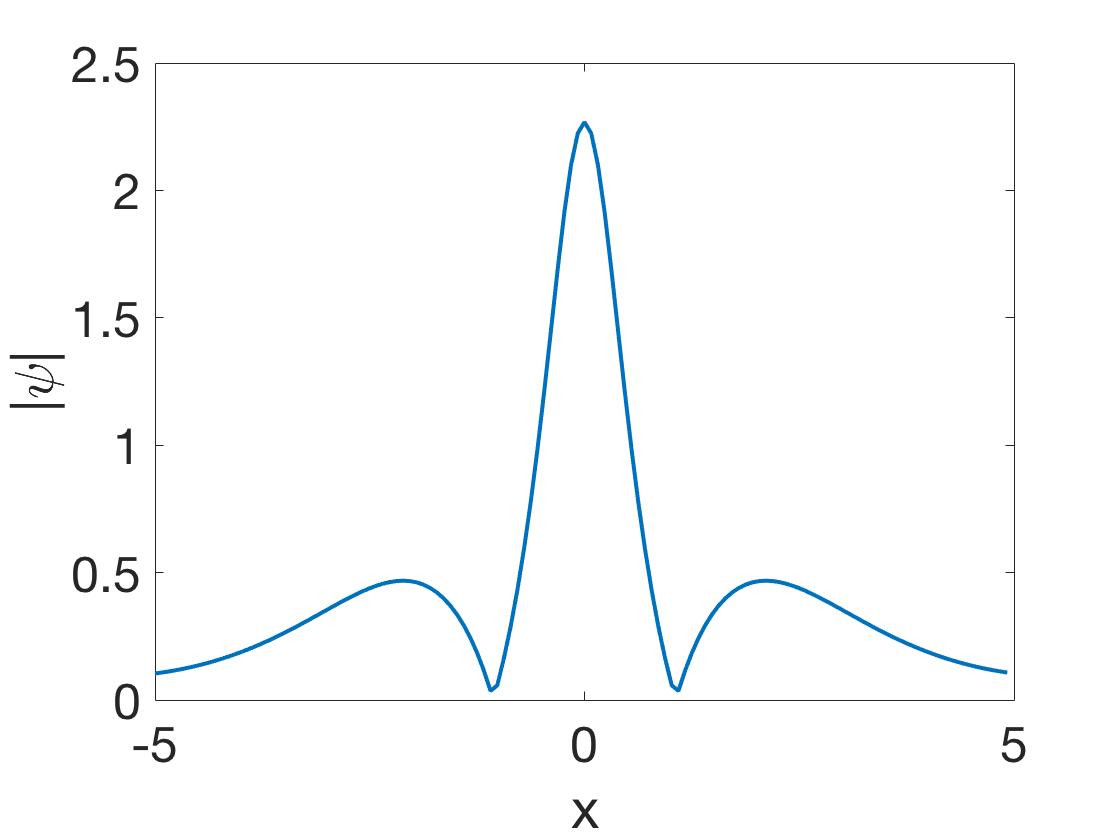}}
	
	\subfloat[]{\includegraphics[width=.25\textwidth]{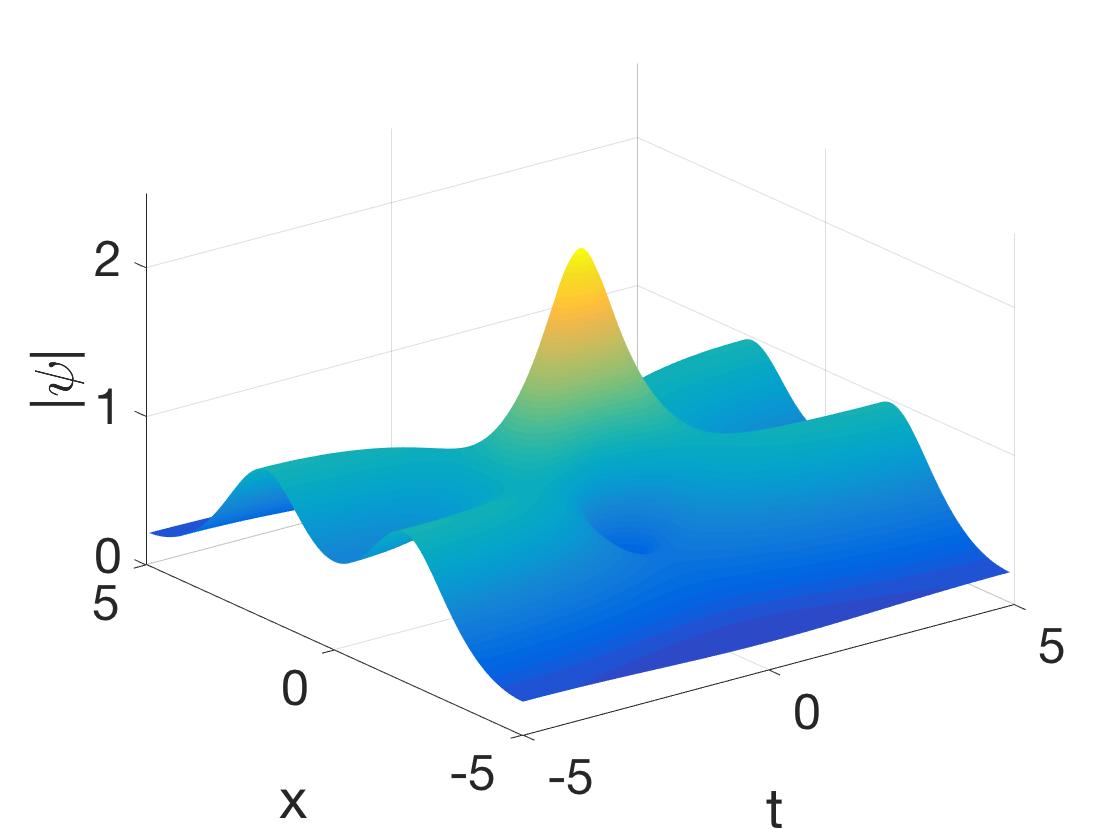}}
	\subfloat[]{\includegraphics[width=.25\textwidth]{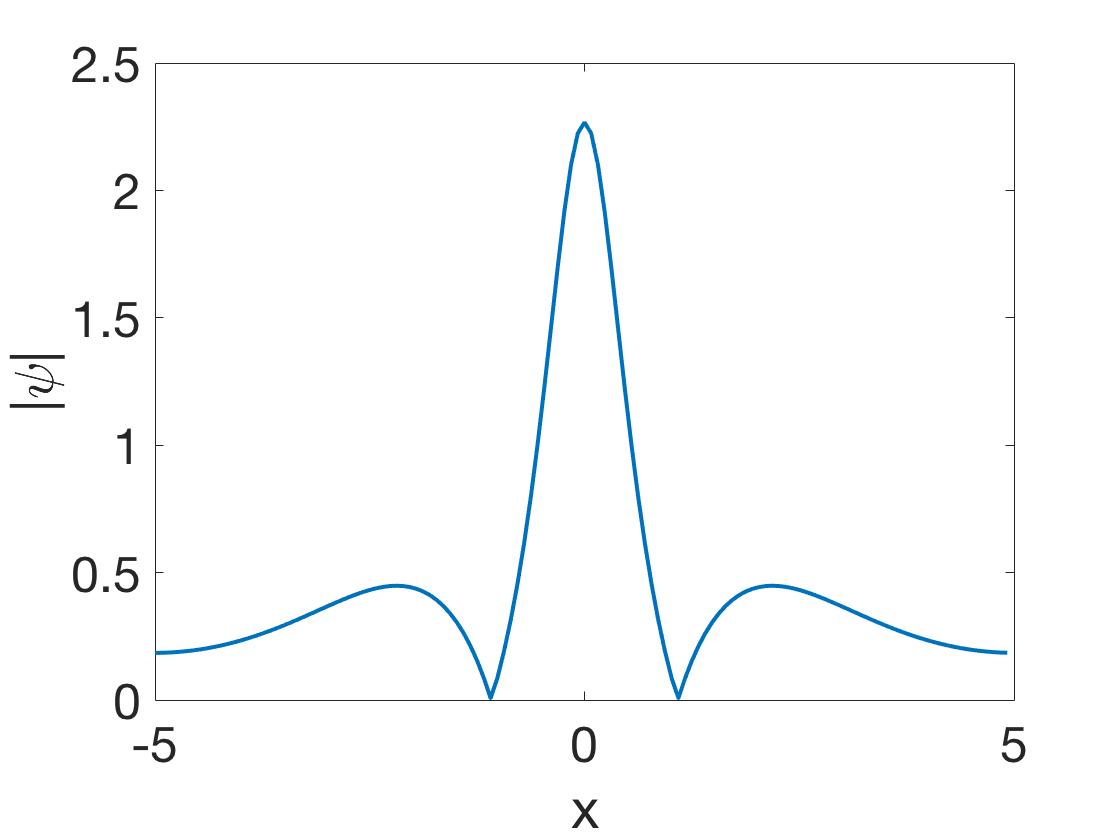}}
     \caption{Comparison of an exact cnoidal rogue (a) with the numerical solution (c). Slice at $t=0$ of the exact solution (b) and numerical solution (d). Again, we see very good agreement between the two solutions.} \label{Fig2}
\end{figure}

As an additional case example for the convergence of the code,
we considered the case of rogue waves
atop an cnoidal (space-time periodic) background~\cite{Ank2} instead of the
usual
constant background.
Fig.~\ref{Fig2}(a) shows an exact, cnoidal rogue wave we obtained by using the procedure (and chosen to be approximately periodic on the domain) of \cite{Ank2}. On the other hand, Fig.~\ref{Fig2}(c) shows the solution obtained by the Newton-CG method. As before, Fig.~\ref{Fig2}(b),(d) show the cross section at time $t=0$ of the exact and numerical solutions, respectively. Again, we see good agreement between the two solutions with only slight differences between the tails.

\begin{figure}[htbp]
	\subfloat[]{\includegraphics[width=.25\textwidth]{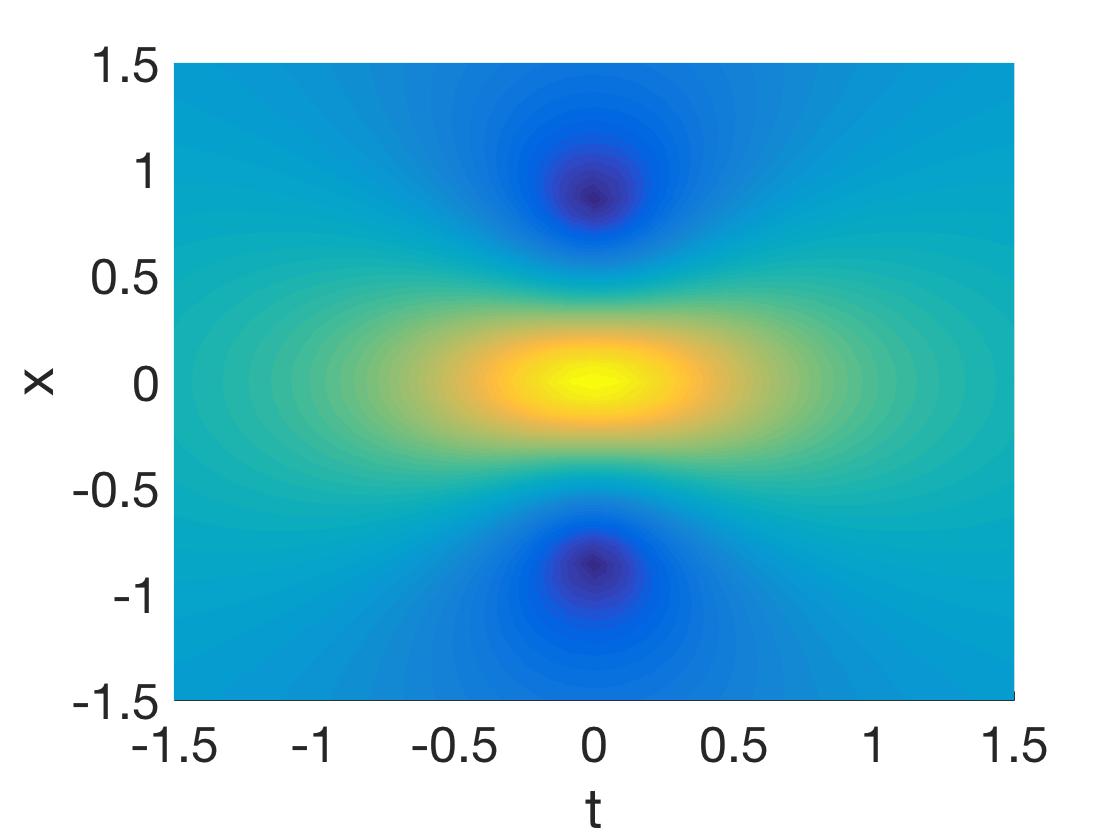}}
	\subfloat[]{\includegraphics[width=.25\textwidth]{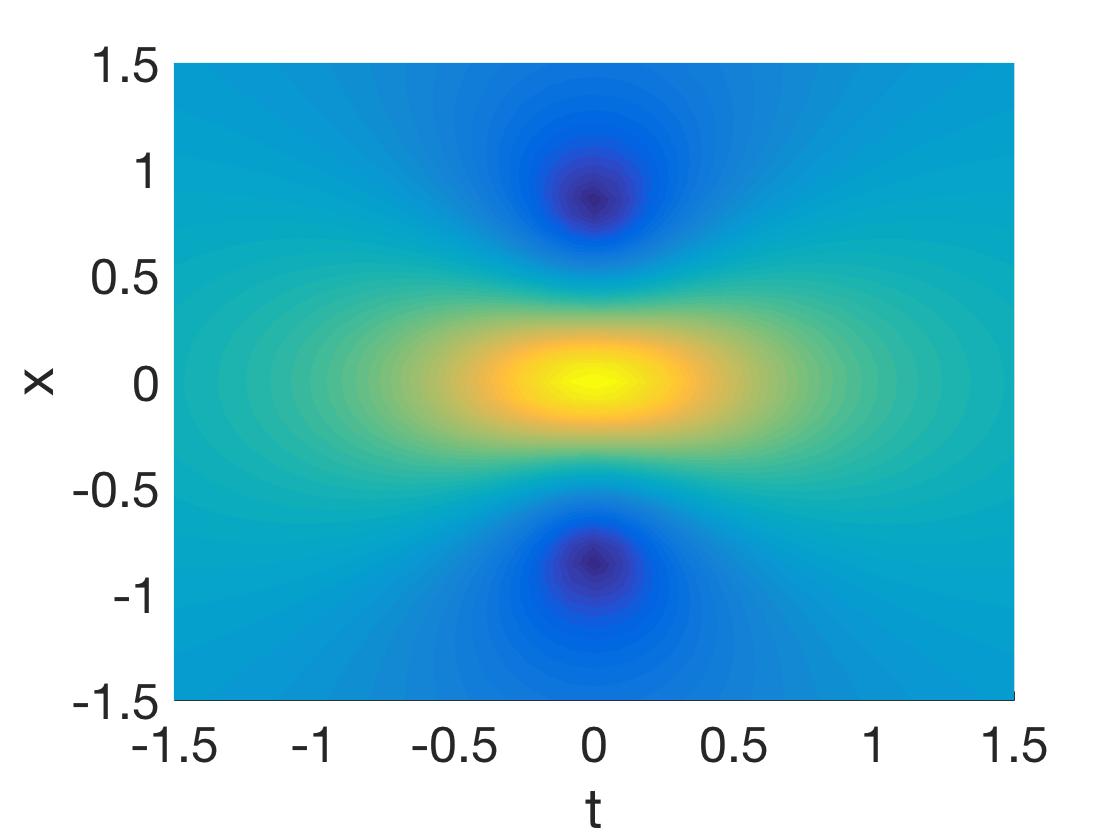}}
	
	\subfloat[]{\begin{tikzpicture}
	\node (tiger) [anchor=south west,inner sep=0] at (0,0) {\includegraphics[width=.25\textwidth]{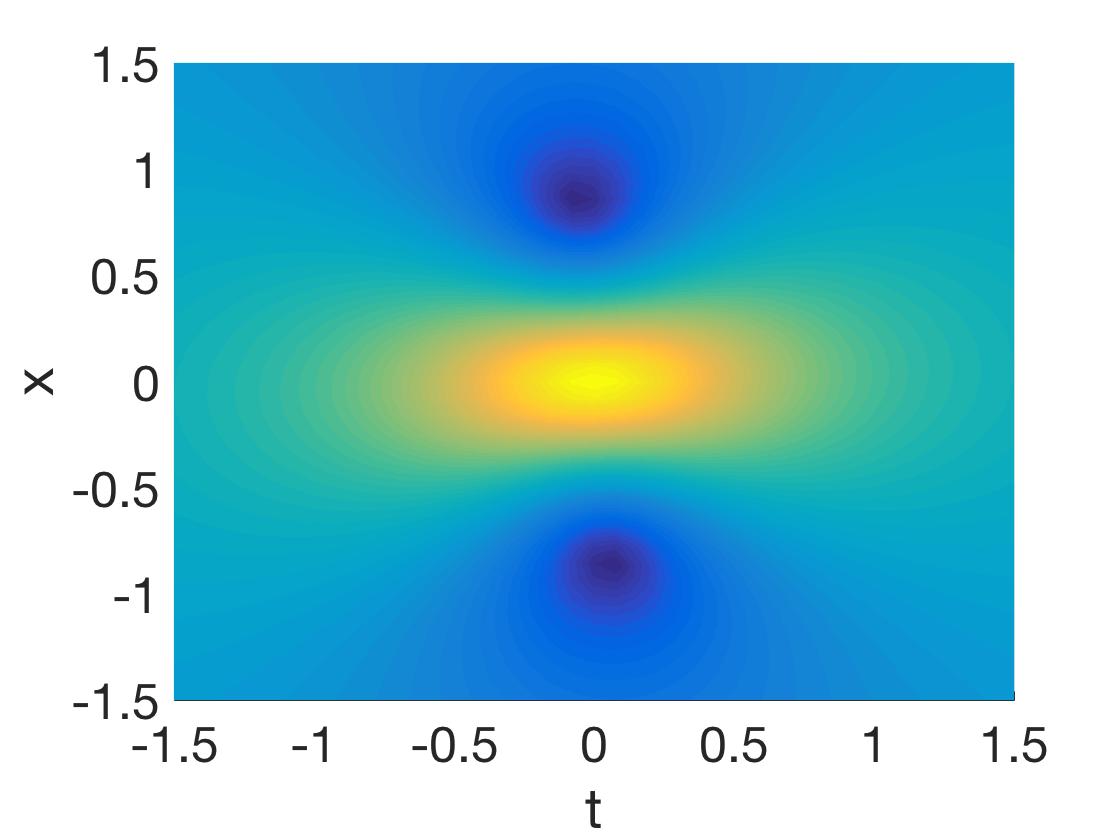}};
        \begin{scope}[x={(tiger.south east)},y={(tiger.north west)}]
          \foreach \i/\j in {{(0.53,.92)/(0.53,.17)}}
            \draw [red,very thick, dashed] \i -- \j;
        \end{scope}
 \end{tikzpicture}}
 \subfloat[]{\begin{tikzpicture}
	\node (tiger) [anchor=south west,inner sep=0] at (0,0) {\includegraphics[width=.25\textwidth]{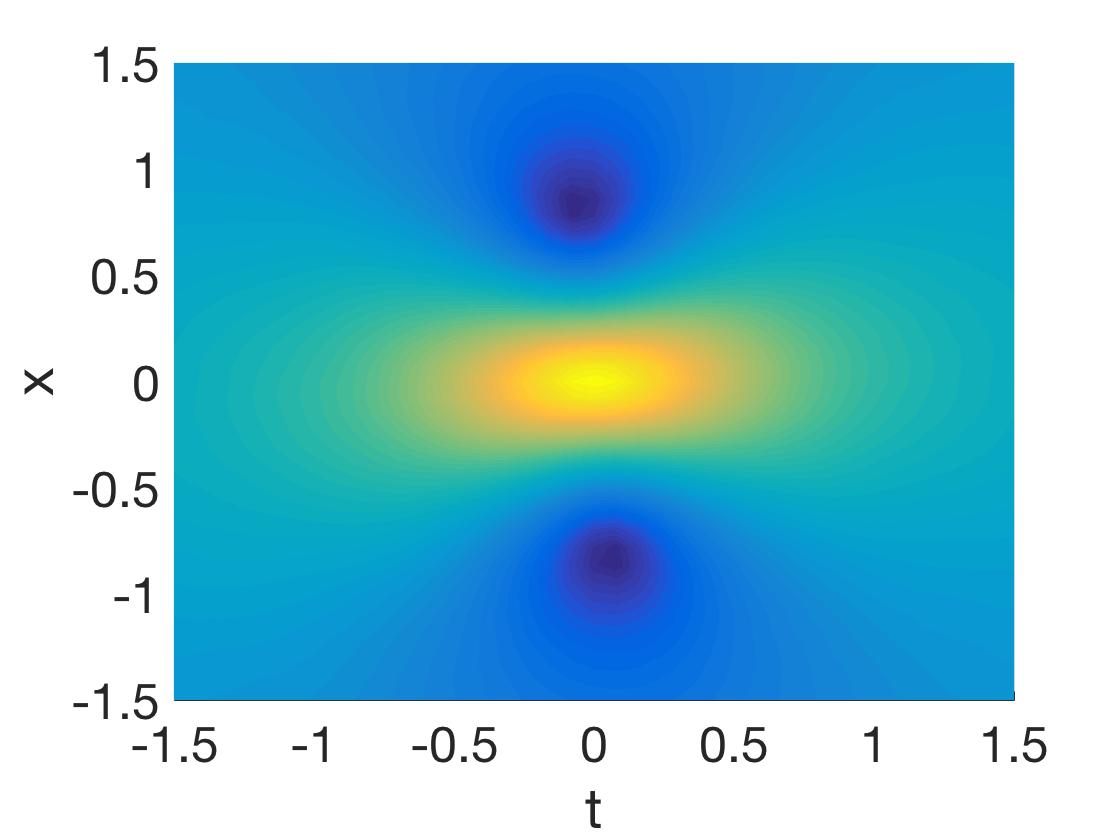}};
        \begin{scope}[x={(tiger.south east)},y={(tiger.north west)}]
          \foreach \i/\j in {{(0.53,.92)/(0.53,.17)}}
            \draw [red, very thick, dashed] \i -- \j;
        \end{scope}
 \end{tikzpicture}}
        \caption{(a) Contour plot of the Peregrine soliton. (b) Corresponding solution obtained numerically. (c) Contour plot of the perturbative solution Eq.~\eqref{eq2}, while (d) is the corresponding numerical solution. Although it is weak,
          there is a discernible asymmetry in panels (c) and (d) (as compared with (a) and (b)) caused by the TOD term. Here we have set $\epsilon=0.02$. We
          have also included the vertical red bar in (c) and (d) so as to
          highlight the slight rotation (in comparison to the
          $\epsilon=0$ case where the peak and dips of the PS are
          are aligned).}\label{Fig3}
\end{figure}
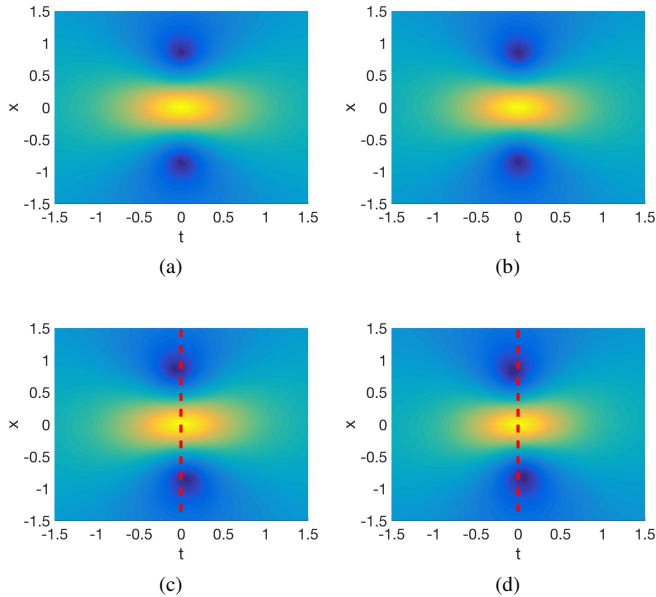

Next, we look for solutions outside of the integrable case scenario
but for which some information is known. Specifically,
in the presence of TOD within Eq.~(\ref{eq1}), the work of~\citep{Ank1} gives a first-order perturbative solution (for $p=1$) in the form
\begin{equation}\label{eq2}
\psi = [\frac{4(1+2it)}{1+4x^2+4t^2} -1+ \frac{i(f-ik)}{(1+4x^2+4t^2)^2}] e^{it}
\end{equation}
where  $f=8x(24  x^2 + 24 t^2 - 6)\epsilon $ and $k= 192 t x\epsilon$. When $\epsilon=0$, this reduces to the standard Peregrine soliton but when $\epsilon \neq 0$ the result is a slightly rotated rogue wave. Fig.~\ref{Fig3}(c) shows the pertubative solution for $\epsilon = 0.02$ while Fig.~\ref{Fig3}(a)
shows the $\epsilon=0$ case for comparison. Although it is
faint, a slight rotation in the counter-clockwise direction can be seen
when comparing the two. On the other hand, Figs.~\ref{Fig3}(b),(d) are the
corresponding solutions that we obtained numerically. Again, good agreement
is found between a predicted
solution (both for $\epsilon=0$ and for $\epsilon \neq 0$)
and our numerical solution. Hence, we confirm that such PS structures
are present in TOD perturbations of the original NLS model.
      
\begin{figure}[htbp]
	\subfloat[$2p=1.8$]{\includegraphics[width=.25\textwidth]{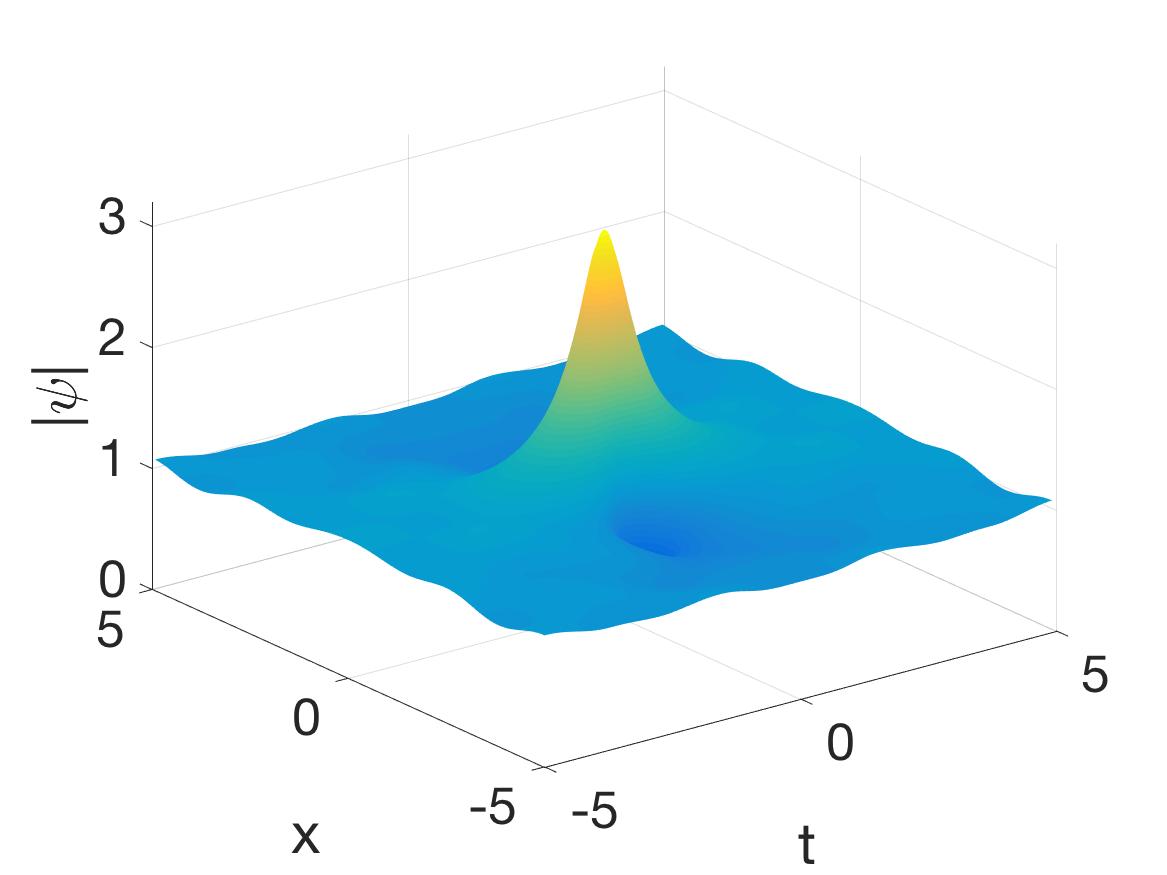}}	
	\subfloat[$2p=1.9$]{\includegraphics[width=.25\textwidth]{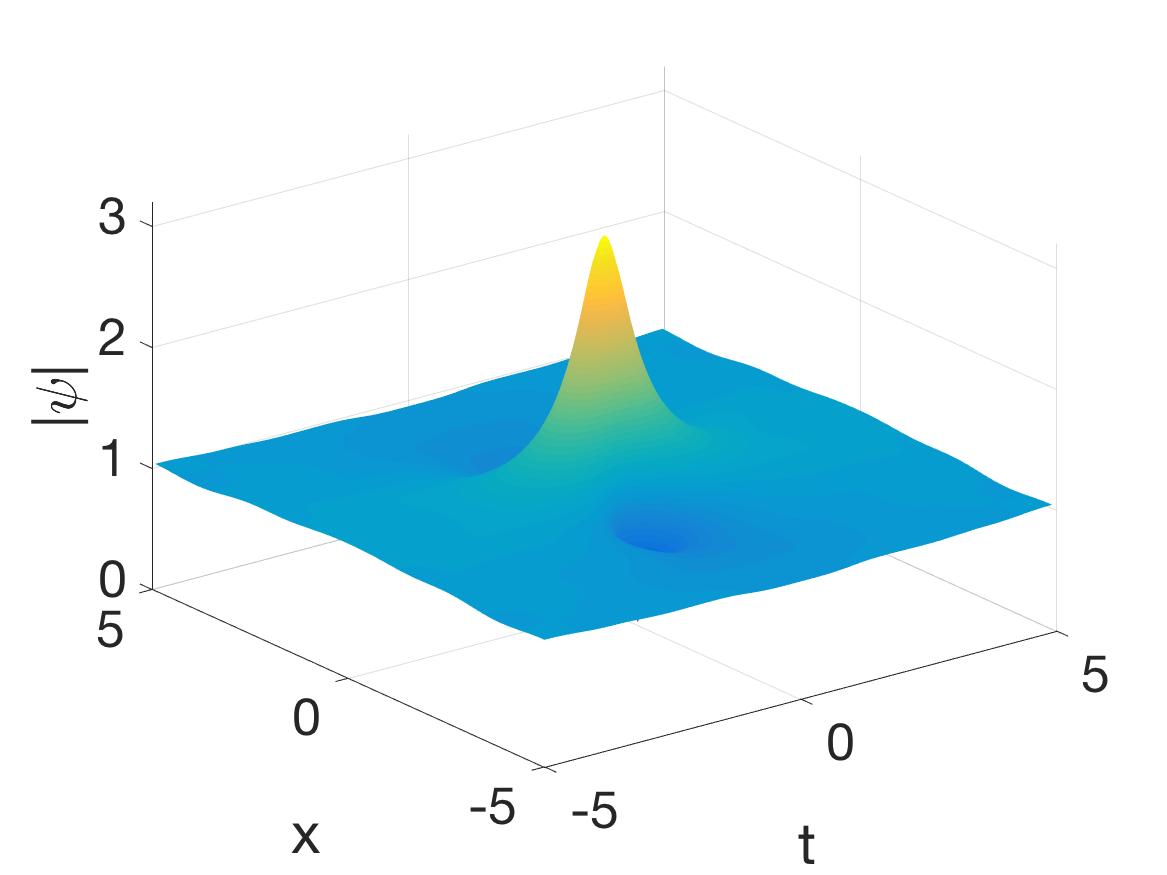}}
	
	\subfloat[$2p=2.1$]{\includegraphics[width=.25\textwidth]{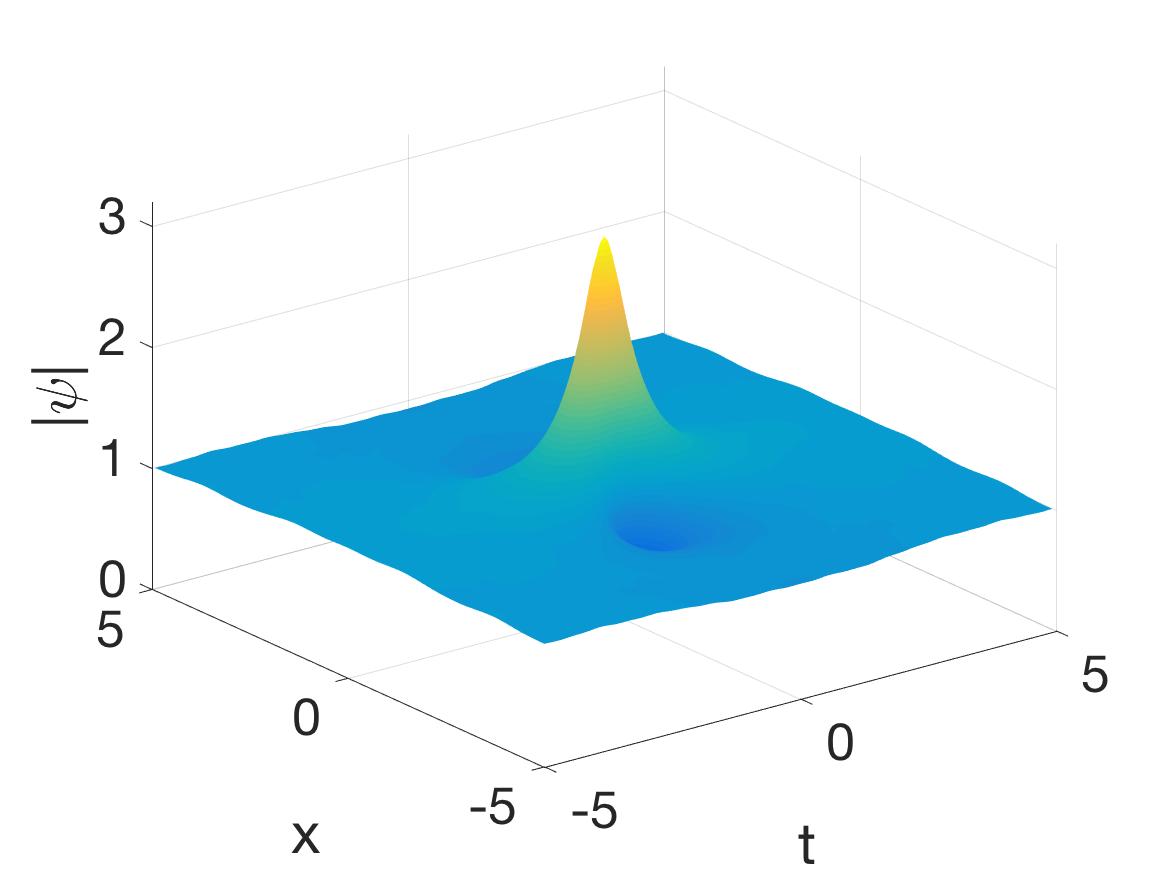}}
	\subfloat[$2p=2.2$]{\includegraphics[width=.25\textwidth]{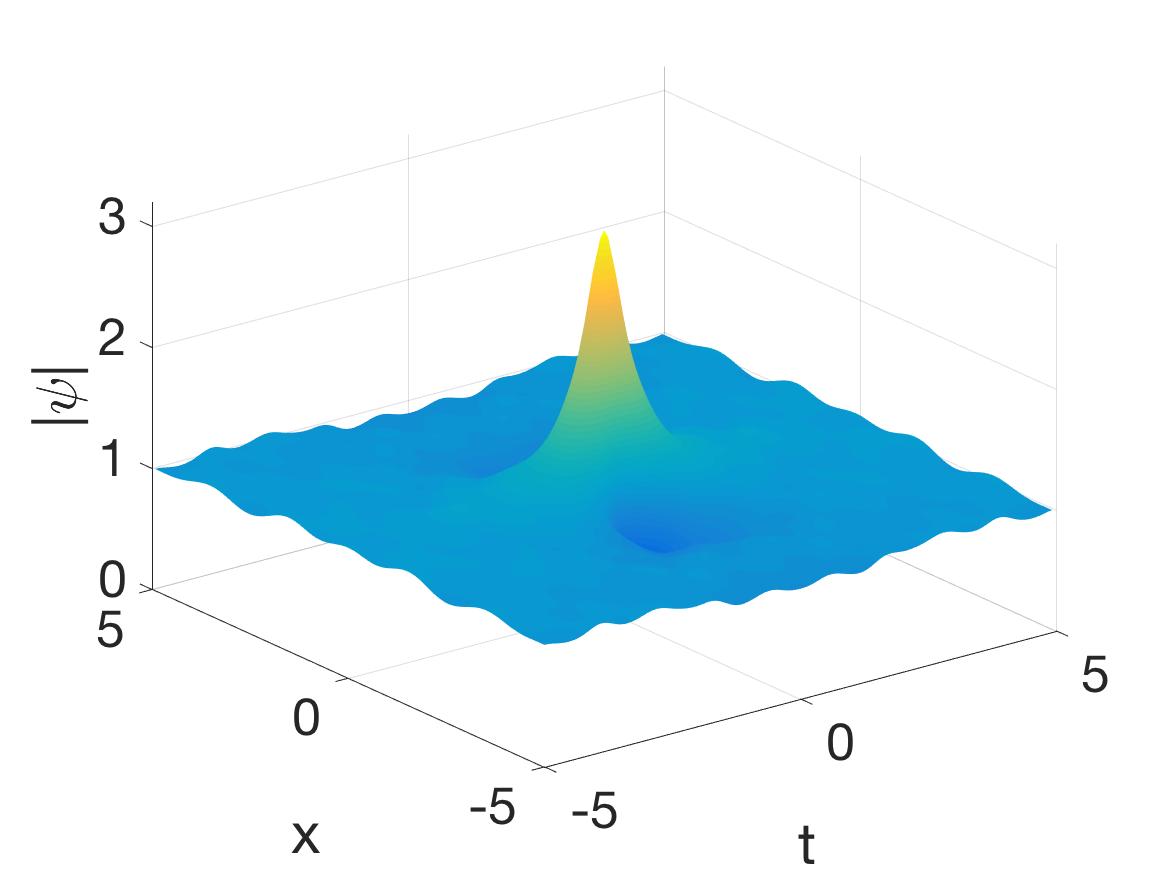}}
        \caption{Family of solutions with $\epsilon=0$ for different
          values of $p$ around the integrable limit of $p=1$. The profiles
        confirm the persistence of the rogue wave.}\label{Fig4}
\end{figure}

Armed with the understanding and expectations suggested by this example,
we now move to a more interesting and unexplored case. In particular,
instead of restraining considerations to the cubic nonlinearity,
we examine general powers $p$, asking whether the rogue wave patterns
will persist. Fig.~\ref{Fig4} shows several solutions obtained via the
Newton-CG method. We initially converged to a solution at
$(p,\epsilon)=(1,0)$ using the PS as our initial iterate. By increasing/decreasing $p$ and using the previously obtained solution as our new initial iterate, we obtained the other solutions via the Newton-CG method. 

For small perturbations in $p$, Fig.~\ref{Fig4}(b),(c), we see that the background remains approximately flat. As far as we know, this is the first
systematic indication that rogue waves exist in the NLS (in
a parametrically continuous way) past the integrable
limit of the cubic nonlinearity; this is a result that
is of particular
importance in connection with the robust experimental observation of such
events. However, for larger perturbations in $p$, Fig.~\ref{Fig4}(a),(d), we see that the background obtains seemingly periodic ripples. One can argue that the issue here is the boundary conditions of the ``true" rogue wave are not
periodic enough to appear in the numerics.
Thus, to make up for this, the system introduces a perturbation on top of the flat background such that the solution will have periodic boundary conditions on the given domain (see also Fig.~\ref{Fig7}(d)).
Irrespective of these non-uniformities
in the background, the presence of a wave that appears out of nowhere
and disappears without a trace is eminently transparent in these
{\it converged} solutions. 

\begin{figure}[htbp]
	\subfloat[$2p=1.8$]{\includegraphics[width=.25\textwidth]{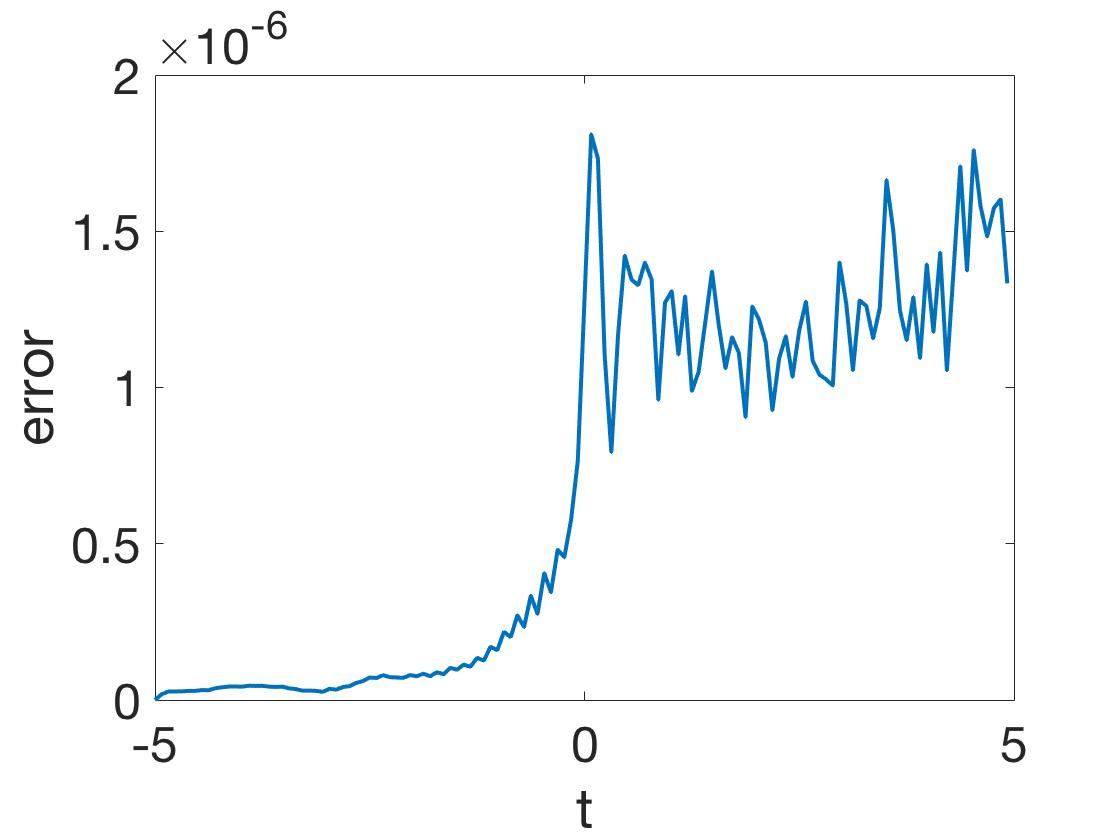}}
	\subfloat[$2p=1.9$]{\includegraphics[width=.25\textwidth]{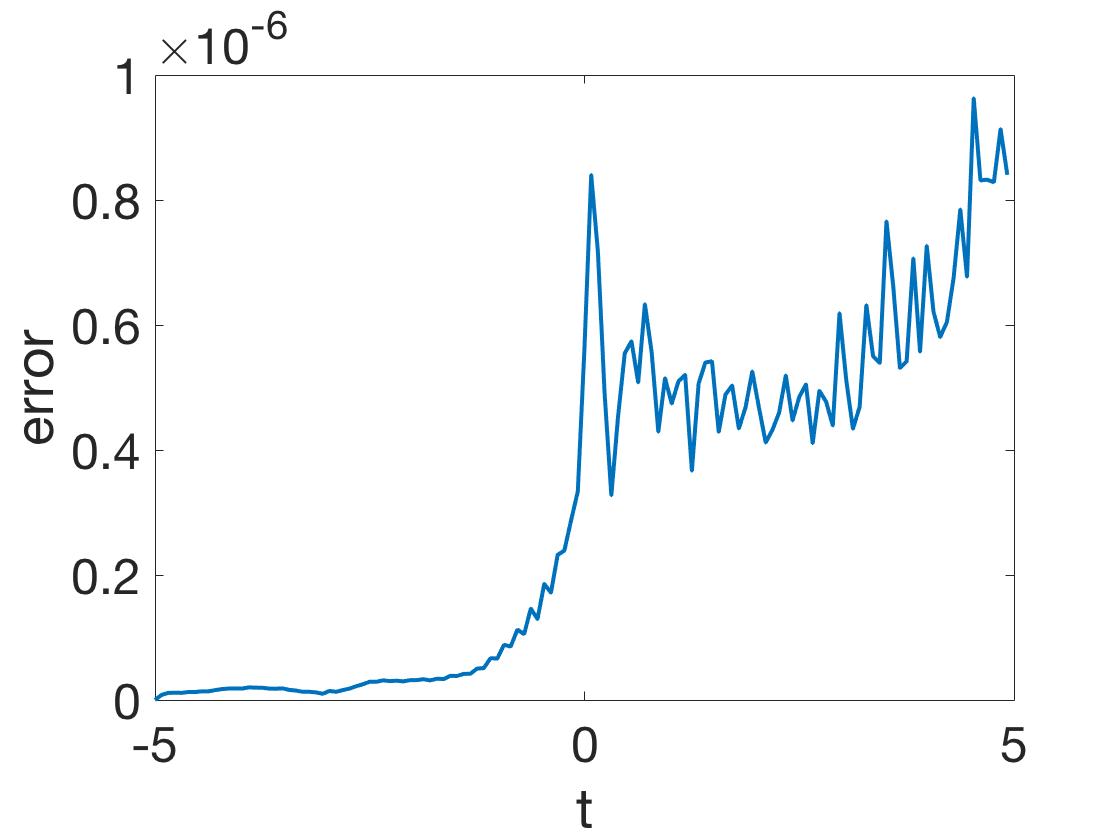}}
	
	\subfloat[$2p=2.1$]{\includegraphics[width=.25\textwidth]{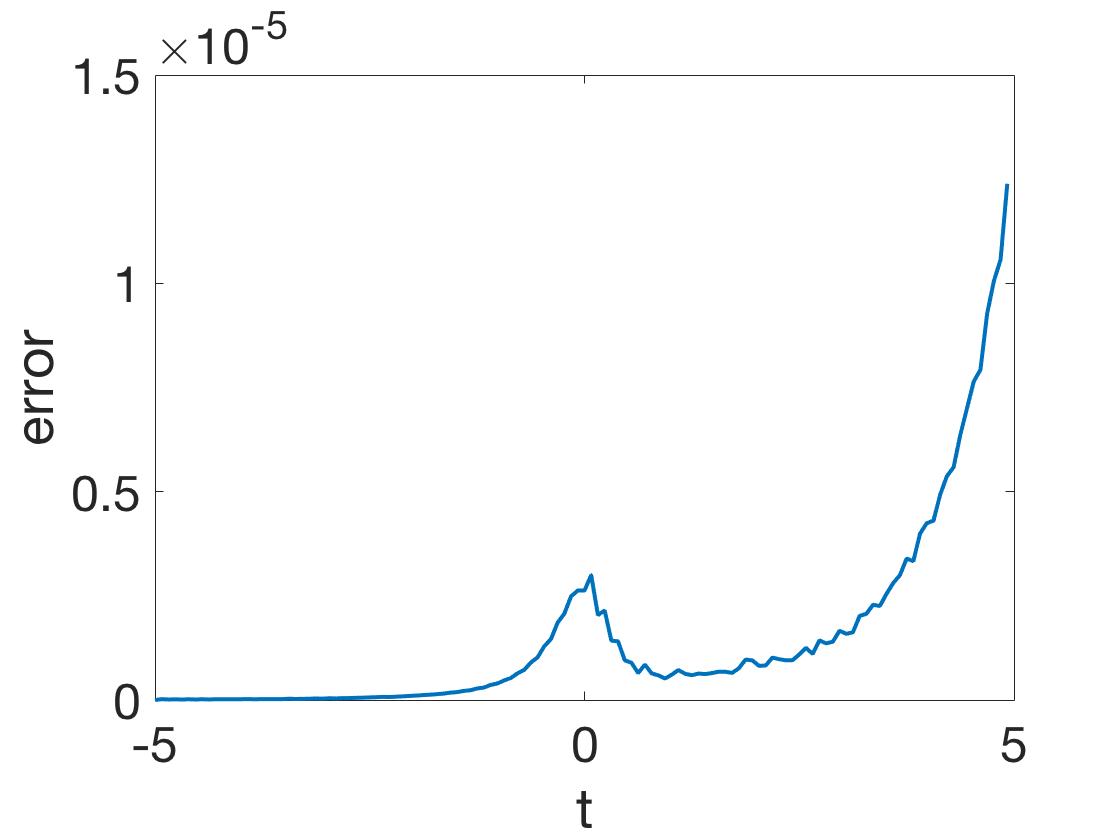}}
	\subfloat[$2p=2.2$]{\includegraphics[width=.25\textwidth]{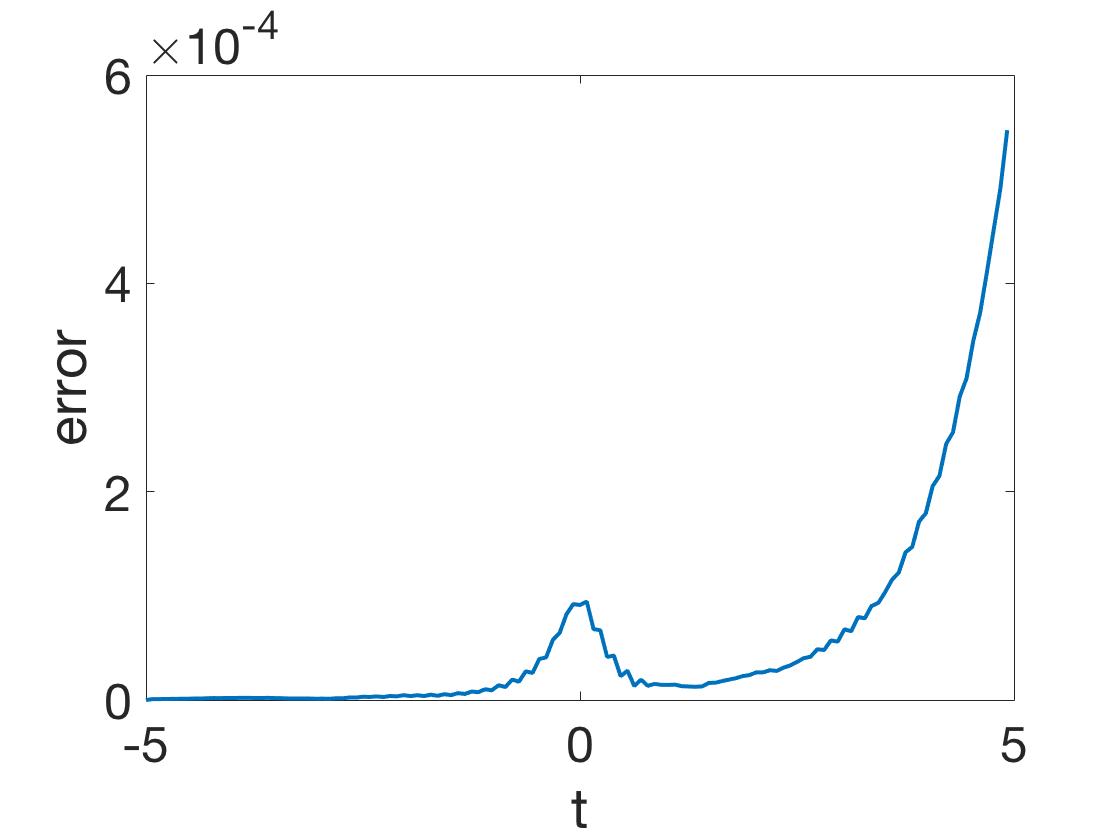}}
        \caption{Direct numerical simulation
          results confirming that evolution through
          the ETDRK4 method yields excellent
          agreement with the Newton-CG findings. The error is measured
          via the $L^{\infty}$ norm (in space) of the difference
          at each point in time between the time-evolved (ETDRK4) and
          the Newton-CG solution. Here $\epsilon=0$.}\label{Fig5}
\end{figure}

It is crucial to note here that to ensure that these are proper solutions
of the original PDE of Eq.~(\ref{eq1}), we have performed direct
numerical simulations with the ETDRK4 time-stepping algorithm \citep{CM,KT}.
Fig.~\ref{Fig5} shows the corresponding results. The general trend in all diagrams is that the solutions agree until slightly after the peak of the rogue wave begins to decay. This is consistent with the occurrence and growth of the modulation instability of the background, further confirming the existence of these objects. 

\begin{figure}[htbp]
	\subfloat[$2p=1.8$]{\includegraphics[width=.25\textwidth]{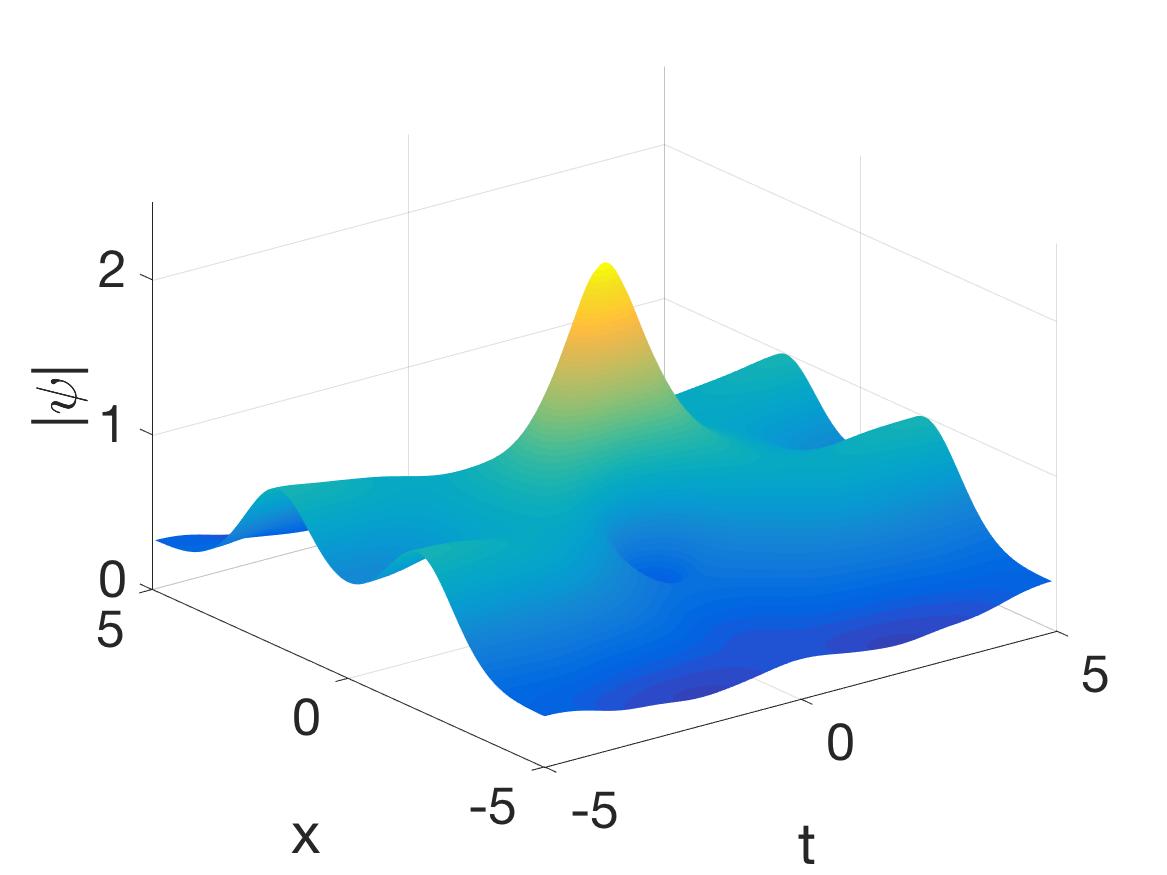}}
	\subfloat[$2p=1.9$]{\includegraphics[width=.25\textwidth]{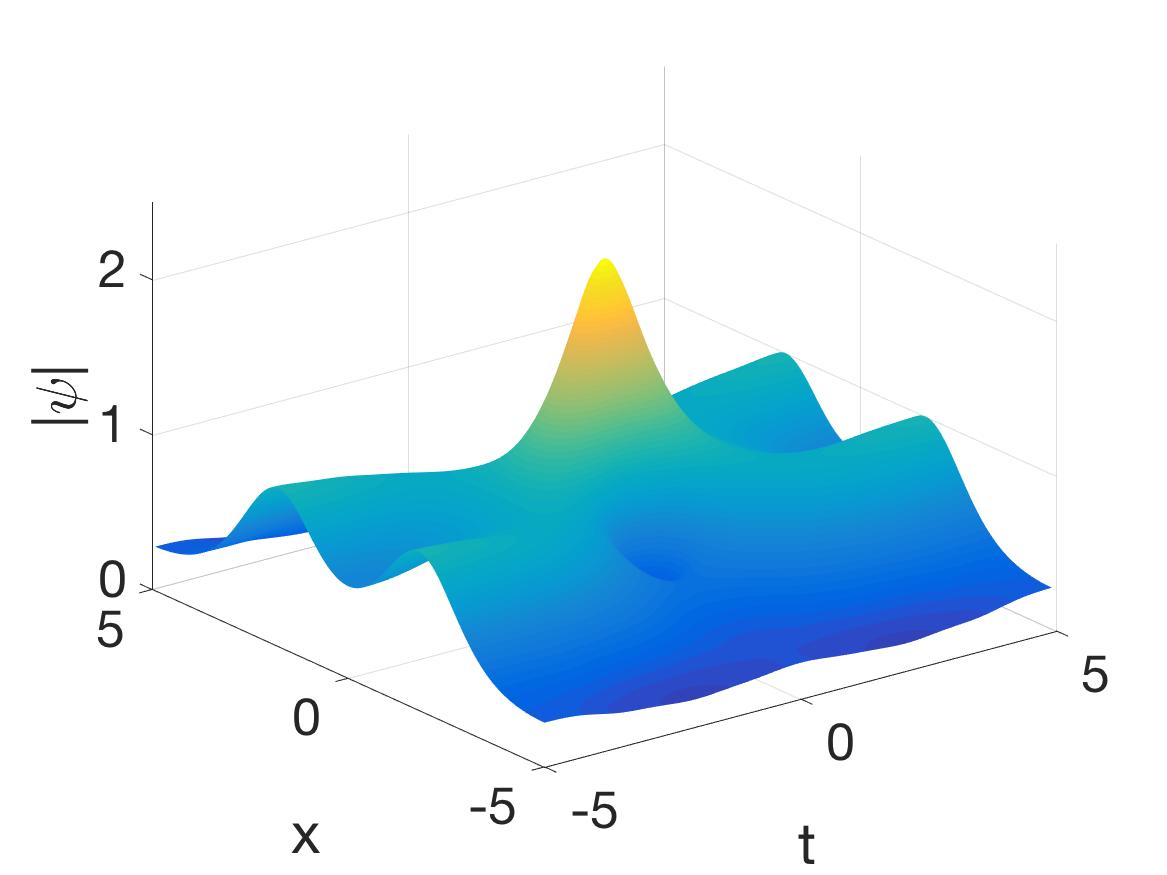}}
	
	\subfloat[$2p=2.1$]{\includegraphics[width=.25\textwidth]{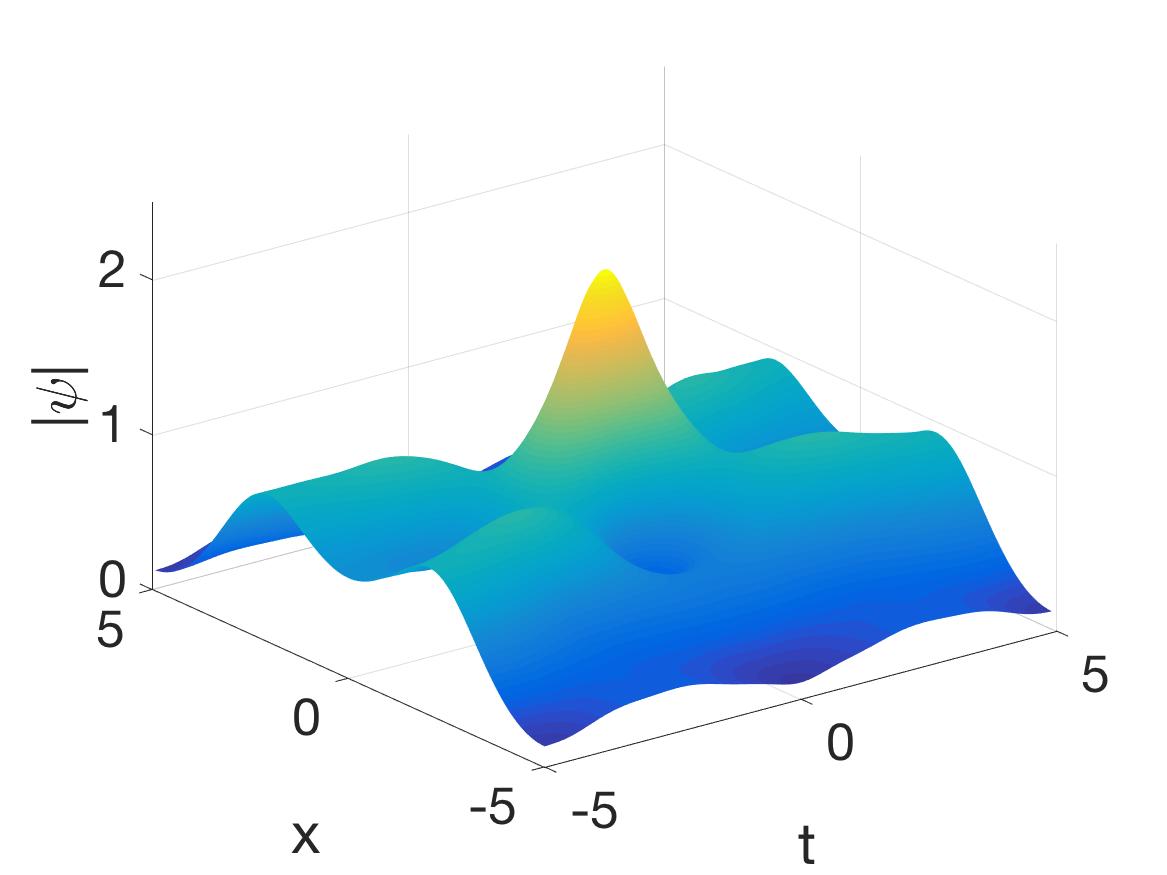}}
	\subfloat[$2p=2.2$]{\includegraphics[width=.25\textwidth]{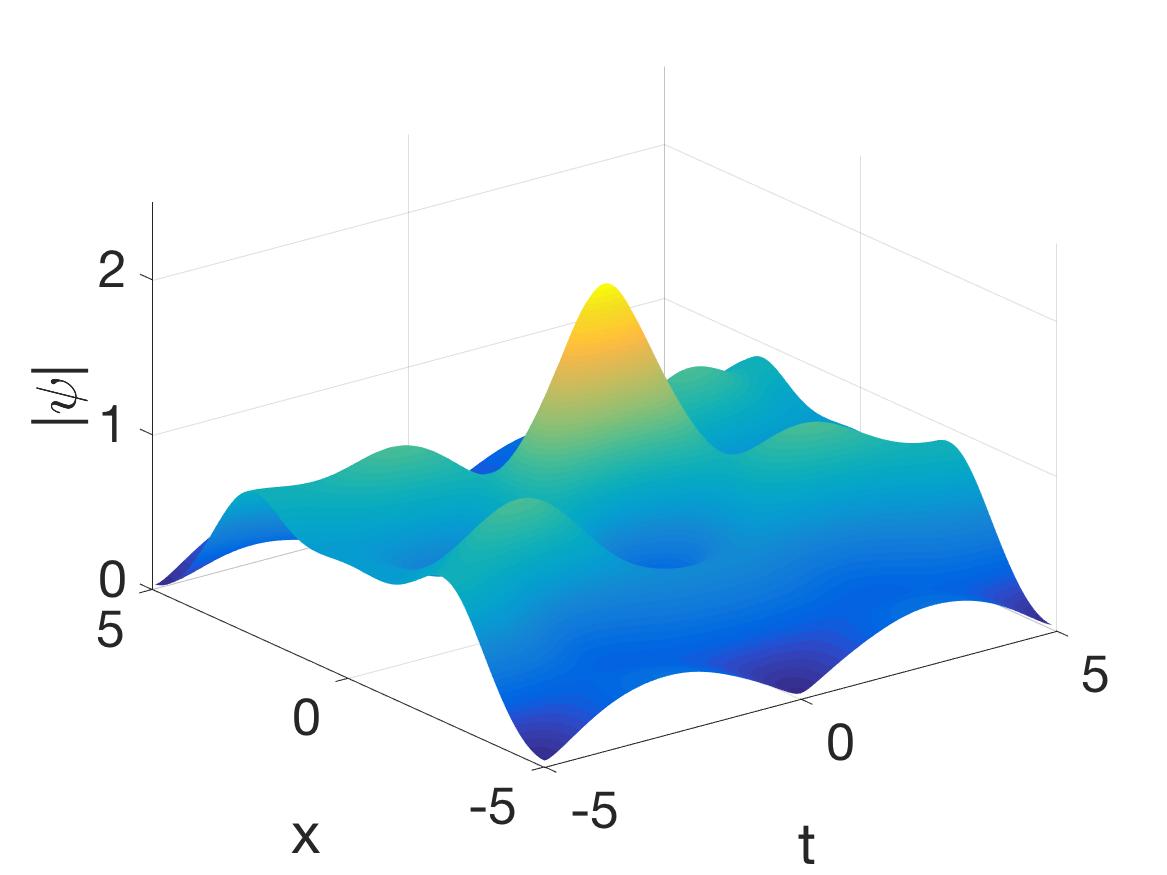}}
        \caption{Same as Fig.~\ref{Fig4} but now for the
          family of rogue waves on top of the cnoidal background with
          $\epsilon=0$. }\label{Fig8}
\end{figure}

To confirm the generality of the findings, 
we have also examined the waveforms on top of the cnoidal background in Fig.~\ref{Fig8}. These have also been identified through
our Newton-CG technique for different values of $p$. 
It is interesting to note that in this case
the system converges to a state without a small periodic perturbation to the background. Furthermore,  we were able to perform the continuation
in this case from $p=0$ all the way to $p=2$. Additionally,
the solutions obtained via the ETDRK4 integrator agree with the solutions obtained by the Newton-CG far better than those obtained on the flat background.
A posteriori, it can be argued that this is to be expected considering these
solutions more naturally conform to the imposed periodic boundary conditions.

Lastly, Fig.~\ref{Fig7} serves
to make the case that the solutions of interest exist
not only along the axes of our two-dimensional $(p,\epsilon)$-plane, but also for nonzero values of both parameters, i.e., under
combinations of different perturbations. Here, we have
verified the convergence of the Newton-CG iterative approach
to a profile bearing a rogue wave for different values of
$p$, and $\epsilon=0.02$.

\begin{figure}[htbp]
	\subfloat[$2p=1.8$]{\includegraphics[width=.25\textwidth]{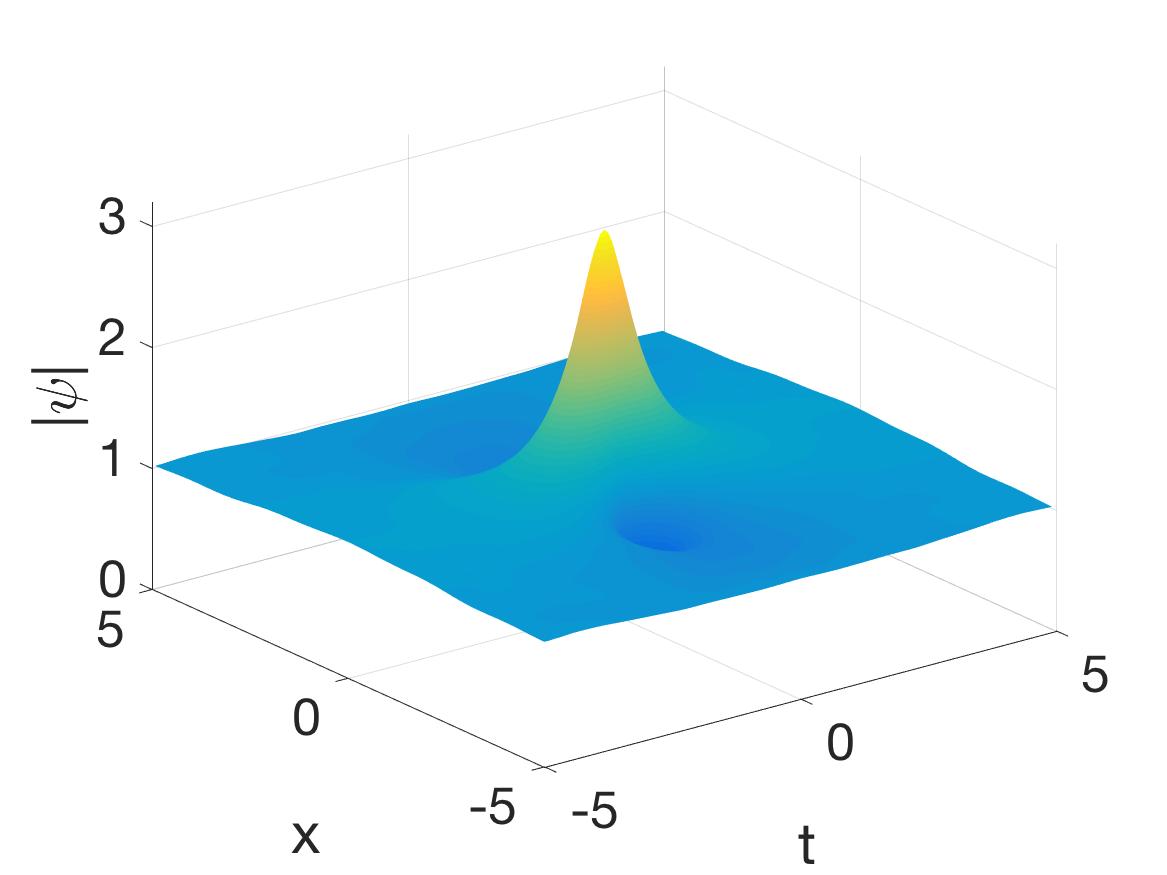}}	
	\subfloat[$2p=1.9$]{\includegraphics[width=.25\textwidth]{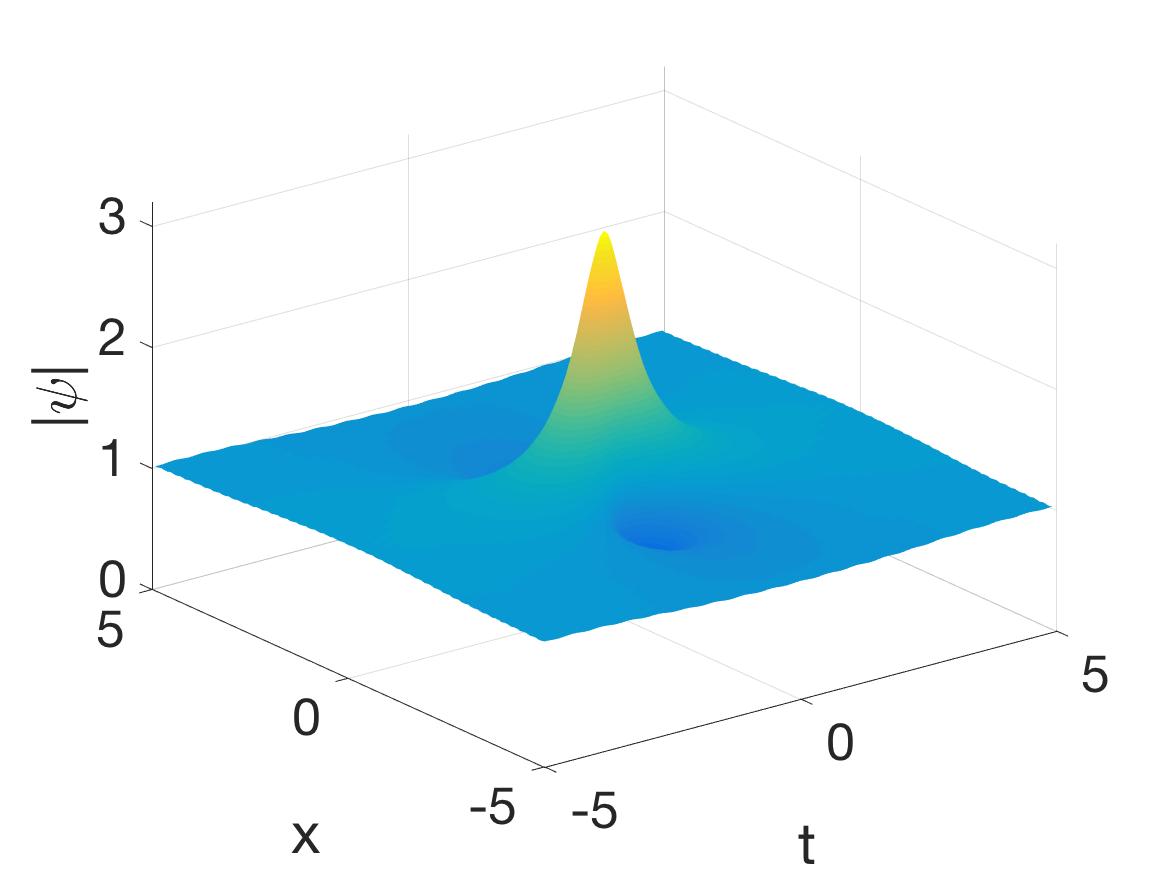}}
	
	\subfloat[$2p=2$]{\includegraphics[width=.25\textwidth]{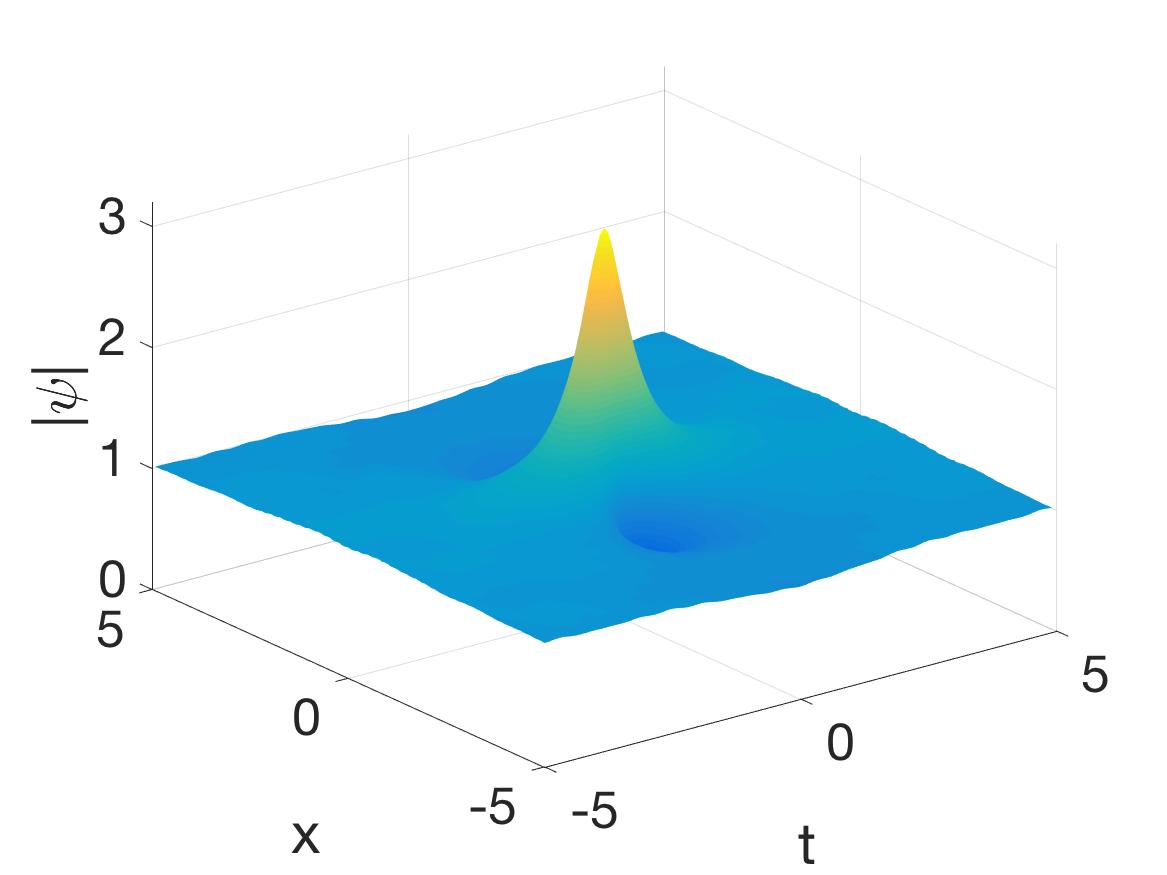}}
	\subfloat[$2p=2.1$]{\includegraphics[width=.25\textwidth]{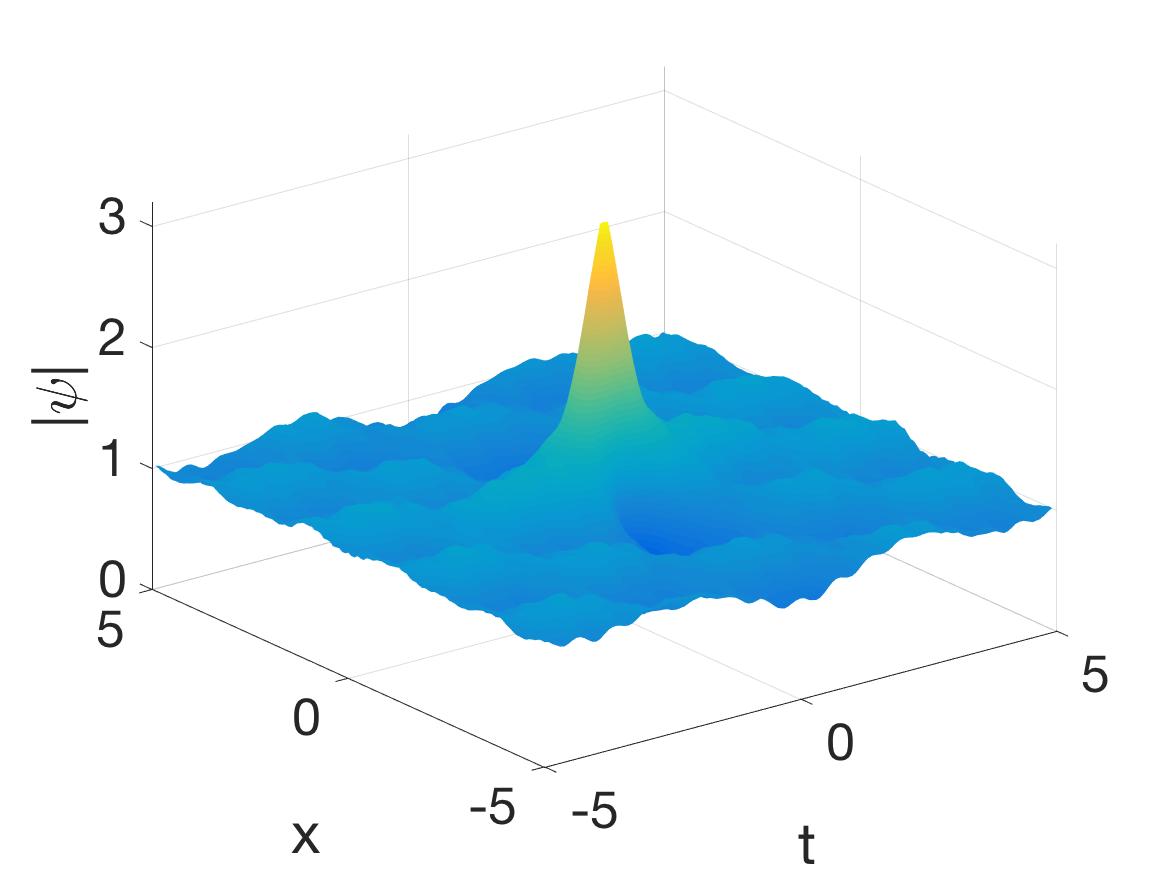}}
        \caption{Persistence of rogue waves for both $p\neq 1$ and $\epsilon \neq 0$
          through the family of solutions with $\epsilon=0.02$.}\label{Fig7}
\end{figure}

\section{Conclusions \& Future Work}

In the present work, we have examined the question of whether rogue
wave structures can persist under different types of perturbations
in the realm of NLS equations. We have adapted a computational
approach based on a Newton-conjugate gradient method to identify the
Peregrine soliton solutions, both on top of a homogeneous background
and on top of a cnoidal-wave background. We have verified that such
solutions can be identified both for the case of a power law
nonlinearity (beyond the cubic integrable limit) and for a
model with third-order dispersion, as well as in a model that
combines both of the above integrability-breaking
perturbations.

These findings pave a new avenue of understanding of such extreme
wave events. They enable us to seek them {\it beyond} the narrow
confines of integrability in a systematic way that does not
need the Lax pair formulation and analytical or perturbative
solutions. At the same time, they suggest numerous questions for
further investigations. Starting with the computations performed,
they enhance (e.g. through the ETDRK4 results confirming the solutions
identified) the belief that all of these solutions are rather
unstable due to their unstable background. Hence, the necessity of
a framework in which these solutions are understood as metastable
and/or present without a homogeneous background is progressively becoming
more dire, so as to justify, among other things, their
undisputed emergence in
experiments and in realistic physical settings.
Perhaps such a framework is that of the gradient catastrophe
of~\cite{berto}, yet this is still a topic worthwhile of further study.

In a different vein, the numerical method used here employs periodic
boundary conditions. As a result the solutions obtained are, effectively,
periodic in space and time. It would certainly be desirable to deploy
a method that either involves the well-known  asymptotics
(in space and time) of the Peregrine soliton, or one that accounts
(in some way reminiscent of transparent boundary conditions)
for the algebraic decay of the wave structure. This is an interesting
direction for further numerical developments. Such studies are presently
under consideration and will be reported in future publications. 

{\it Acknowledgements.} PGK gratefully acknowledges discussions
with and inspiration from
J. Cuevas-Maraver, D. Frantzeskakis, N. Karachalios, G. James
and M. Haragus on the topic of rogue waves. P.G.K. also
acknowledges that this paper was made possible by NPRP
grant \# [8−-764−-1−-160] from the Qatar National
Research Fund (a member of Qatar Foundation).
The findings achieved herein are solely the responsibility of the
authors.

\section{Supplementary Information}
\subsection{Methods}
To obtain the results in this paper, our primary tool was the Newton conjugate gradient method, suitably adapted, of the earlier work \cite{Yang}.
The method approximates a solution of a partial differential
equation (PDE) by expanding it into complex exponentials and then
solving the resulting system for the amplitudes (i.e.,
a psuedospectral Galerkin method). Finding a solution is accomplished with Newton's method except that the linear system is
solved iteratively via the conjugate gradient method. Two major benefits of
this method is that it is spectrally accurate
in both space and time,
and it is relatively straightforward to code (see the following section).

%(see the Supplementary Information (SI) for an example). 
%This is relavant for publication. Don't delete.

However, due to the choice of basis functions, the method implicitly assumes periodic boundary conditions.
In that light, we attempt
to use a domain that is sufficiently large for the rogue wave
structures to approach their equilibrium state, yet small enough
to avoid issues with either the size of the computation or the
instability of the background. Thus, the computations reported
herein have been performed in a space-time domain $[-5,5] \times [-5,5]$.
Lastly, we terminated the Newton-CG
iterations once the $L^\infty$ error in the residual drops
below $10^{-8}$.

To confirm dynamically the results of the Newton-CG iteration, we also used time-integration techniques. Specifically, we used the fourth order in time, spectral in space, time integrator ETDRK4 \citep{CM,KT}. For the initial condition in the integrator, we used the $t=-5$ time slice (of the relavant solution) obtained from the Newton-CG method. We then integrated this out to $t=5$. As mentioned above, this choice of time domain allows for the comparison of the two solutions \emph{before} modulation instability enters significantly into the dynamics.

\subsection{Newton-CG code for the NLS Peregrine Soliton}
The following is an example of the code we used to obtain the rogue waves appearing in this paper. It is a suitable adaptation to
the present setting of the code of~\cite{Yang}.
\begin{figure}
\begin{Verbatim}
%Newton-CG for finding Rogue Waves in NLS with TOD (see epsn below)
% iu_t + (1/2)u_xx - i\epsilon u_xxx + |u|^{2p} u - u =0
%Note that we've factored out the time-dependent phase factor,
%resulting in the additional -u term in the equation.

clear
%SET UP
Lt=10; Lx=10; 
Nt=2^7; Nx=2^7;                           %Number of points/basis functions
t=-Lt/2:Lt/Nt:Lt/2-Lt/Nt; x=-Lx/2:Lx/Nx:Lx/2-Lx/Nx;                %tx-grid
kt=[0:Nt/2-1 -Nt/2:-1]*2*pi/Lt; kx=[0:Nx/2-1 -Nx/2:-1]*2*pi/Lx;
[T,X]=meshgrid(t,x); [KT,KX]=meshgrid(kt,kx);

errormax=1e-8;                                             %Stop Condition.
errorCG=1e-2;          %Controls how accurately you solve the linear system
                       %per Newton iteration. Varying this can help/hurt
                       %the total iteration count.

%LINEAR (DERIVATIVES) PART
c=5;                          %Used in the preconditioner
p=1; epsn=0;                       %Nonlinearity and \psi_xxx coeffecient
K2=KT+(1/2)*KX.^2 + epsn*KX.^3;                               %Wave Numbers 
fftM=c+K2.^2;                        %Preconditioner. Since K2 can be zero,
                                     %this choice makes the preconditioner
                                     %symmetric positive definite.

%INITIAL CONDITION
PP=@(t,x) 1- 4*(1+2i.*t)./(1+4*x.^2 + 4*t.^2);
U=PP(T,X);


ncg=0;                                       %Number of CG-iterations used.
ITER=20000;                         %Total number of CG-iterations allowed.
flag=1;                                                   %While Condition.
while flag==1  && ncg <= ITER
    
  %Error and Stop Condition  
  L0U=ifft2(-K2.*fft2(U))+ (abs(U).^(2*p)).*U - U;                %Equation
  err=max(max(abs(L0U)))                                             %Error

  if err < errormax                                         %Stop Condition
     flag=0;
  end

              %Directional derivative of the nonlinear term (since it isn't
  
                                                                
  dN= @(D) (p+1).*(abs(U).^(2*p)).*D +...
                                   (p).*(U.^2).*(abs(U).^(2*p-2)).*conj(D);
  
  L1= @(D) ifft2(-K2.*fft2(D))-D + dN(D);              %Linearized Equation          
  L1A= @(D) ifft2(-K2.*fft2(D))-D + dN(D);                %Adjoint Equation
  
              %Note: If F=0 is the equation, then this code actually solves
              %the linear system:     DF^* DF du = - DF^* F
              %where "^*" means adjoint. The reason is that DF^* DF
              %should ideally be symmetric, positive-definite. 
              

  %CG-Iteration
  R=-L1A(L0U);                         
  
  DU=0*U;
  MinvR=ifft2(fft2(R)./fftM);
  R2new=sum(sum(conj(R).*MinvR));
  R20=R2new;
  P=MinvR;
  while(abs(R2new) > abs(R20)*errorCG^2 && flag==1) 
    L1P=L1(P); LP=L1A(L1P);
    a=R2new/sum(sum(real(conj(P).*LP)));
    DU=DU+a*P;
    R=R-a*LP; MinvR=ifft2(fft2(R)./fftM);
    R2old=R2new;
    R2new=sum(sum(real(conj(R).*MinvR)));
    b=R2new/R2old;
    P=MinvR+b*P;
    ncg=ncg+1;  
  end
  
  %Newton Iteration
  U=U+DU;
end

figure; surf(t,x,abs(U))
shading interp, lighting phong
xlabel('t'); ylabel('x'); zlabel('abs(U)');
\end{Verbatim}
\end{figure}
\end{document}